\documentstyle[]{mn}
\input psfig.sty
\def\diff{{\rm d}}
\def\tot{{\rm tot}}
\def\eff{{\rm eff}}
\def\DV{de Vaucouleurs{}}
\def\Se{S\'ersic{}}
\def\arcsec{\hbox{$^{\prime\prime}$}}
\def\mumean{\langle \mu \rangle}
\def\FP{Fundamental Plane{}}
\title[The specific entropy of elliptical galaxies]%
{The specific entropy of elliptical galaxies: an explanation for profile-shape
distance indicators?}

\author[G.B. Lima Neto et al.]
{G.B. Lima Neto,$^{1,2}$ D. Gerbal,$^{1,3}$ I. M\'arquez$^{4}$\\
$^1$ Institut d'Astrophysique de Paris, CNRS, Universit\'e Pierre et
Marie Curie, 98bis Bd Arago, F-75014 Paris, France\\
$^{2}$ Instituto Astron\^omico e Geof\'{\i}sico/Universidade de S\~ao Paulo, 
Av. Miguel Stefano, 4200, 04301-904 S\~ao Paulo/SP, Brazil\\
$^3$ DAEC, Observatoire de Paris, Universit\'e Paris VII, CNRS (UA 173),
F-92195 Meudon Cedex, France\\
$^4$ Instituto de Astrof\'{\i}sica de Andaluc\'{\i}a,
CSIC, Apdo 3004, 18080 Granada, Spain\\
e-mail:~gastao@iap.fr, gerbal@iap.fr, isabel@iaa.es\\
}

\date{Accepted ???. Received ????; in original form ????}
\pagerange{\pageref{firstpage}--\pageref{lastpage}}
\pubyear{1999}

\begin{document}
\maketitle
\label{firstpage}
\begin{abstract}
Dynamical systems in equilibrium have a stationary entropy; we suggest
that elliptical galaxies, as stellar systems in a stage of
quasi-equilibrium, may have in principle a unique specific entropy.
This uniqueness, \textit{a priori} unknown, should be reflected in
correlations between the fundamental parameters describing the mass
(light) distribution in galaxies. Following recent photometrical work on
elliptical galaxies (Caon et al. 1993; Graham \& Colless 1997; Prugniel \&
Simien
1997), we use the \Se\ law to
describe the light profile and an analytical approximation to its
three dimensional deprojection. The specific entropy is then calculated
supposing that the
galaxy behaves as a spherical, isotropic, one-component system in hydrostatic
equilibrium, obeying the ideal gas state equations.
We predict a relation between the 3 parameters of
the \Se\ law linked to the specific entropy, defining a surface in the
parameter space, an `Entropic Plane', by analogy with the well-known
Fundamental Plane. We have analysed elliptical galaxies
in two rich clusters of galaxies (Coma and ABCG~85) and a group of
galaxies (associated with NGC~4839, near Coma). We show that, for a
given cluster, the galaxies follow closely a relation predicted by the
constant specific entropy hypothesis with a typical dispersion (one standard
deviation) of 9.5\% around the mean value of the specific entropy.
Moreover, assuming that the specific entropy is also the same for
galaxies of different clusters, we are able to derive \textit{relative}
distances between Coma, ABGC~85, and the group of NGC~4839. If the
errors are only due to the determination of the specific entropy (about 10\%), then
the error in the relative distance determination should be less than
$20\%$ for rich clusters. We suggest that the unique specific entropy may
provide a physical explanation for the distance indicators based on the
\Se\ profile put forward by Young \& Currie (1994, 1995) and recently
discussed by Binggeli \& Jerjen (1998).
\end{abstract}


\section{Introduction}\label{intro}

Elliptical galaxies are believed to be self-gravitating systems in a
quasi-equilibrium state. They are far from being the simple, dull
objects they were thought to be until the early 70's: high spatial and
spectroscopic observations have revealed that elliptical galaxies show
a great variety of fine structures (e.g. Kormendy 1984, Bender \&
M\"ollenhoff 1987). However, underlying these fine structures,
elliptical galaxies have a striking regularity concerning their radial
light distribution. Indeed, until the mid 80's, it seemed that the
brightness profile of elliptical galaxies could be well described by
the de Vaucouleurs (1948) $R^{1/4}$ law (e.g. Kormendy 1977).
Nevertheless, systematic departures from the $R^{1/4}$ law were observed
(Michard 1985; Schombert 1986, 1988) and, more recently, thanks to
higher quality observations, it was shown (Caon et al. 1993; Graham \&
Colless 1997) that elliptical galaxies in a very wide range of sizes
can be well modelled by the \Se\ (1968) empirical law (expressed in
magnitude):
\begin{equation}
 \mu =  \mu_{0}+ k_{n}(\frac{R}{R_{\eff}})^{(1/n)}
\label{sersic0}
\end{equation}
a non-homologous generalization of the $R^{1/4}$ law.
 $\mu_0$ is the central brightness, $R_{\eff}$ is the effective radius and
$k_{n}$
a function depending on $n$ (see below, \S\ref{generalites}).

Besides the light profile, another very important regularity observed in
elliptical galaxies is the so-called Fundamental Plane (Dressler et al. 1987b;
Djorgovski \& Davis 1987). Elliptical galaxies populate a two dimensional
surface in the three dimensional space defined by the effective
radius, $R_\eff$, the mean surface brightness inside this radius,
$\mumean_\eff$, and the central line of sight velocity dispersion, $\sigma_0$.
The understanding of such a relation should provide important clues to the
formation and evolution of these galaxies.

What is the origin of the relations among the parameters describing the light
(and mass) distributions of elliptical galaxies? If these objects are, as we
believe, in equilibrium, then the virial theorem should apply and we can derive
for a given mass distribution a relation among the gravitational radius, total
mass, and mean velocity dispersion. This relation can be `translated' in the
observed \FP\ with the help of some hypotheses related to the mass to light
ratio,
light distribution, and hydrostatic equilibrium (Bender et al. 1992, Ciotti \& 
Lanzoni 1997). The
validity of these hypotheses is, however, still a matter of debate (cf. for
instance Prugniel \& Simien 1997).

Other than its theoretical importance, the scaling relations of the
fundamental parameters of galaxies may have a practical application as a
distance indicator. Thus, like the \FP\ (Djorgovski \& Davis 1987), the
$D_n$--$\sigma$ relation (Dressler et al. 1987a),
Tully \& Fisher (1977) method for spiral galaxies, etc.,
the parameters of the \Se\ law can also be used to
derive distances (Young \& Currie 1994, 1995).

In this paper we will look for fundamental relations among the parameters
describing the light distribution of elliptical galaxies.  The starting
point is
also the fact that these galaxies are in equilibrium, but we will use the
maximum entropy postulate of thermodynamics instead of the virial theorem.  We
will derive an expression relating the free parameters of the \Se\ profile to
the specific entropy.  We will also give a number of useful and simple
expressions to calculate the astrophysical quantities related to the \Se\
profile and its three dimensional deprojection.  We will then show that the
specific entropy relation holds for at least two clusters of galaxies and we
suggest that it may, in principle, be used as a relative distance indicator like the shape-profile distance indicators introduced by Young \& Currie (1994, 1995).

\section{The entropy}\label{entropie}
\subsection{Thermodynamics}\label{thermo}

The regularity observed in the light distribution of elliptical galaxies
may be
conveniently understood as an evolutionary convergence. Independently of the
initial conditions, these galaxies evolve towards a single class of objects.
Since the stellar two-body relaxation time scale is far greater than the
age of
the Universe, the mechanism behind this evolution is probably the so-called
violent relaxation (Lynden-Bell 1967). The rapidly changing global
gravitational
potential (due to an initial collapse or merging) produces a mixing in phase
space within the same time scale as the collapse or merging itself.

The second postulate of thermodynamics says that, when a dynamical system
is in equilibrium, then this system is in a state of maximum entropy. If, for a
given system, there exists only one state of thermodynamic equilibrium
that can be reached, independently of the initial conditions, then it
is associated with a stable state of single, global maximum entropy.
Gravitational systems, however, present some difficulties: given an
arbitrary isolated, self-gravitating system of point particles, the
entropy can grow indefinitely by collapsing part of the mass in the
centre and expanding the outer parts to infinity, while still
conserving the total mass and energy. This is the gravo-thermal
catastrophe which can actually occur because a gravitational system
has negative specific heat (Lynden-Bell \& Wood 1968; Saslaw 1985).

Thus, if we try to maximise the entropy (requiring that the mass and energy be 
constant) in order to get the equilibrium (and hopefully stable) configuration, 
we obtain an isothermal sphere (Lynden-Bell \& Wood 1968), which does not match 
the observed elliptical galaxies. A lot of work has been done these last three 
decades to obtain the structure of self-gravitating systems from the maximum 
entropy principle (e.g. Antonov 1962; Lynden-Bell 1967; Shu 1978; Binney 1982; 
Tremaine et al. 1986; Dejonghe 1987; Richstone \& Tremaine 1988; Merritt et al. 
1989; Spergel \& Hernquist 1992; Hjorth \& Madsen 1995). A local maximum entropy 
can be found if some \textit{ad hoc} constraints are imposed (e.g. imposing a 
power law density profile, White \& Narayan 1987).

Here, we adopt a different approach: we just assume the existence of a local 
maximum entropy associated with the elliptical galaxies, \textit{without} trying 
do obtain this entropy from first principles. Except for interacting galaxies, 
it is reasonable to suppose that elliptical galaxies are, at least, objects in 
thermodynamic quasi-equilibrium (the best evidence being the regularity in the 
luminosity distribution). Hence, we will proceed computing the entropy of a 
self-gravitating system using classical thermodynamics. Let us assume that the 
stars in a galaxy behave like an isolated, self-gravitating gas of equal mass 
particles. Consequently, we start by taking the differential of the entropy:
\begin{equation}
\diff S = \frac{1}{T} \diff E + \frac{P}{T} \diff V - \frac{\mu_{q}}{T}
\diff N\, ,
\label{euler}
\end{equation}
where $S, E, V, N, T, P, \mu_{q}$ are, respectively, the entropy, energy, 
volume, number of particles, temperature, pressure, and chemical potential. 
Formula~(\ref{euler}) is nothing more than the fundamental equation of 
thermodynamics (also know as the Euler relation) in the entropy representation 
(e.g. Callen 1960), which applies to any thermodynamic system in equilibrium. 
For an isolated, self-gravitating system, the extensive and intensive parameters 
in Eq.~(\ref{euler}) are usually interpreted using their classical meaning.
For instance, the 
chemical potential $\mu_{q}$ (that appears already in the work of Lynden-Bell 
1967 in the context of stellar dynamics) gives the variation rate of the energy 
with the number of particles, $\diff E/\diff N$, for constant entropy and volume 
(see also Saslaw 1985) or, from the statistical physics point of view, $\mu_{q}$ is 
related to the Fermi energy of the system (Stiavelli 1998). The temperature is 
defined in terms of the velocity dispersion $\sigma$, as $kT = m\sigma^2$ ($m$ 
is the mean particle mass of the system).

We now need the equations of state in order to
describe the properties of the `gas' of stars. There is no known exact equation
of state for a gas of self-gravitating particles, so we will use the
ideal gas equations which are usually a good approximation for hot,
structureless, rarefied gases:
\begin{eqnarray}
E &=& {3\over 2} N\; kT\; ; \quad PV = N\; kT\; ; \nonumber \\
\mu_{q} &=& kT \ln \left(
{{N \over V}{\left[{{h^2 \over 2\pi m\; kT}}\right]}^{3/2}}
\right) \, ,
\label{etats}
\end{eqnarray}
where $k$ and $h$ are the Boltzmann and Planck constants.
Although the above equations of
state are a good approximation for \textit{non-interacting}, mono-atomic, hot
gas (in fact, these equations violate the third law of thermodynamics or
Nernst
postulate, i.e. that the $S\rightarrow 0$ as $T\rightarrow 0$), their
application for interacting particles is doubtful. Modifications of the
Eqs.~(\ref{etats}) have been proposed; Bonnor (1956, and references therein)
discusses a modification of Boyle's Law that takes self-gravity into account.
However, that modification was derived for a bounded, isothermal sphere and it
adds considerable complication to the problem without really putting away the
simplifying hypotheses. Therefore, we take a pragmatic approach and we will
use
the simplest physically acceptable equations of state, Eqs.~(\ref{etats}).
The validity or not of our
approach should be verified \textit{a posteriori}.

Thus, we can now write the entropy as a function of the mass density $\rho
= m N/V$
and pressure $P$. Using Eq.~(\ref{euler}) and Eqs.~(\ref{etats}), we
obtain, after straightforward algebra:
\begin{equation}
{m \over k}{\diff S \over \diff V} = {5 \over 2}\rho + \rho
\ln\left({\rho^{-5/2}\; P^{3/2} \left({{2\pi \over h^2}}\right) m^4}\right)\, .
\label{diffEntro}
\end{equation}
We now define the specific entropy $s \equiv S/M$, where $M$ is the
total mass of the system. Therefore we write:
\begin{equation}
s = {1 \over M}\int_V^{} \ln\left({ \rho^{-5/2}\; P^{3/2}}\right)\rho\, \diff V
 + \mbox{constant} \, ,
\label{specificEntro}
\end{equation}
where the `constant' above does not depend on the density
distribution and we will drop it hereafter. Notice that the above equation is
the same used by White \& Narayan (1987).
Equation (\ref{specificEntro}) can be written in a more compact form
using the definition of a mean value, $\langle x\rangle = \int x f(x) \diff x
/ \int f(x) \diff x$, and interpreting the density $\rho$ as a distribution
function:
\begin{equation}
s = \left\langle{\ln\left({ \rho^{-5/2}\; P^{3/2}}\right)}\right\rangle\, .
\label{meanThermoEntro}
\end{equation}
In Appendix~1 we show that we can obtain the same form for the
specific entropy as in Eq.~(\ref{meanThermoEntro}) using the Boltzmann-Gibbs
definition of the entropy.

\subsection{Entropy and light profile}\label{2.2}

The parameters needed for the evaluation of the specific entropy of early
type galaxies are not accessible by direct observation. The calculation of the
specific entropy must thus be obtained using the available light profile
and some additional hypotheses.

It is clear that, to calculate the mass
distribution needed to compute the specific entropy of the galaxy
from the observed light distribution, we will
need to introduce a mass-to-luminosity ratio, $M/L$.
In general, the radial variation of the $M/L$ ratio in elliptical galaxies
is poorly known, basically because its determination depends on knowledge of
the intrinsic shape and detailed dynamics of the galaxy. So far, there is
little evidence for radial gradients of $M/L$ up to $2R_\eff$ (de Zeeuw
1992, Bertin et al. 1994, Carollo et al. 1995, Rix et al. 1997).

Therefore, we suppose that the galaxy behaves like a one-component system. This
means that either (I) the stellar component is dominant (the invisible
matter is negligible), or (II) the luminous and dark matters are
distributed in the same manner, or else (III) the dark matter has a
very extended distribution, being flat and less important than the
visible component in the central region (it is the case if the
luminous part of an elliptical galaxy is embedded in a very large dark
matter halo).

The velocity dispersion tensor is not \textit{a priori} isotropic in a
galaxy, so that the radial and transverse velocity dispersions may not
always be equal (for a discussion see Binney \& Tremaine 1987).
Moreover, normal elliptical galaxies are known to have little
rotation (cf. de Zeeuw \& Franx 1991 and references therein).
We suppose therefore, when using the equations of state~(\ref{etats}),
that the gas of stars is actually
isotropic and with null angular momentum. This simplification will
enable us to calculate the
pressure of the gas of stars without adding any other parameter to the problem.

The next step for really computing the entropy is to determine the radial mass
distribution of galaxies.

\section{The S\'ersic law}
\subsection{Generalities}\label{generalites}

A generalization of the \DV\ profile has been proposed by \Se\ (1968), 
Eq.~(\ref{sersic0}). His profile has been widely used in recent years for 
modelling and theoretical studies of elliptical galaxies (e.g., 
Ciotti 1991; Caon et al. 1993; Colless \& Dunn 1996; Courteau et al. 
1996; Prugniel \& Simien 1997; Graham et al. 1996; Graham \& Colless 1997). The 
\Se\ law is one of the most compact and simple expressions that describes very 
precisely the light distribution in a wide range of early-type galaxies.

Instead of the historical form of the \Se\ law, Eq.~(\ref{sersic0}),
we will use throughout this paper the following mathematically
more compact and natural form:
\begin{equation}
 \Sigma(R) =  \Sigma_{0} \exp(-(R/a)^{\nu})\, .
\label{sersic}
\end{equation}
Sometimes we call it the $\nu$-model, by analogy to the $\beta$-model
(very often used to describe the x-ray profiles of galaxy clusters).

The \Se\ profile is characterised by a normalisation parameter, $\Sigma_{0}$; a 
scale length, $a$; and a structural (or shape) parameter, $\nu$ ($\nu \equiv 
1/n$). The \DV\ profile is of course equivalent to $\nu = 0.25$. We call $\nu$, 
$a$, and $\Sigma_0$ the \textit{primary parameters} of the \Se\ law.

From the mathematical point of view, the \Se\ profile primary parameters are 
naturally independent, i.e. if -- as we will show later -- after a fit, some 
correlation among these parameters occurs, it will be only due to physical 
properties. This is not the case when other kinds of parameters such as 
effective radius or effective brightness are used, because they involve 
combinations (usually non linear) of the primary parameters (cf. Gerbal et al. 
1997 and the end of \S\ref{Correlations}). Such parameters we call
the \textit{secondary parameters} of the \Se\ law.

We will now give some quantities of astrophysical interest and/or useful for
the present paper. Calculations are generally straightforward or are
presented in
the Appendix~2. For the sake of simplicity, we will often give approximate but
analytical expressions for these quantities. The analytical approximations
are obtained
by fitting convenient simple functions to the exact values computed
numerically. We will refer to the accuracy of the analytical approximations of
a given quantity $\zeta$
in terms of $(\zeta_{\rm theo} - \zeta_{\rm app})/\zeta_{\rm theo}$,
where $\zeta_{\rm theo}$ is the exact theoretical value
and $\zeta_{\rm app}$ is the analytical approximation.
Notice that we do not use the calculations
provided by Ciotti (1991), since he based his calculations on a numerical
deprojection of the \Se\ law while we use an analytical approximation
described
below.

The total luminosity inside the radius $R$ is readily obtained by
integrating the profile given in Eq.~(\ref{sersic}):
\begin{equation}
L(R)=\frac{2\pi a^{2}}{\nu}\; \Sigma_{0} \; \hbox{\Large$\gamma$} \!
\left(\frac{2}{\nu},\left(\frac{R}{a}\right)^{\nu}\right)\, ,
\label{Ltotsersic}
\end{equation}
where $\gamma(b, x)$ is the standard incomplete gamma function. The total
luminosity is, then,
\begin{equation}
L_{\tot} = a^{2}\; \Sigma_{0} \; L^*(\nu) \, ,
\label{Ltot}
\end{equation}
where $L^*(\nu) \equiv 2 \pi\Gamma(2/\nu)/\nu$ and $\Gamma(x)$ is the complete
gamma function.

The total magnitude, $m_\tot = -2.5\log L_\tot + $ constant, can
be obtained accurately (better than 0.5\%) for $0.1 \leq \nu \leq 2.0$ by
the following analytical approximation:
\begin{equation}
m^*(\nu) = -0.304 \nu - 1.708 \nu^{-1.44} \, ,
\label{mag_tot}
\end{equation}
with $m^*(\nu) \equiv -2.5\log L^*(\nu)$.

The effective brightness, $\mu_\eff$, is the brightness at the
effective radius, $R_\eff$ (defined as the radius having half
the total luminosity inside it), i.e. $\mu_\eff = \mu(R_\eff)$,
and is given by (within 0.5\% accuracy):
\begin{equation}
\mu_\eff= -2.5 \log \Sigma _{0} + 1.0857 (R_\eff/a)^\nu\, .
\label{mueff}
\end{equation}
The mean surface brightness inside $R_\eff$ can be written, by definition, as:
\begin{equation}
\mumean_\eff = m^* + 5\log(R_{\eff}/a) - 2.5\log\Sigma_0 + 2.5\log(2\pi)\, .
\label{mu_eff}
\end{equation}

The effective radius, $R_{\eff}$, is obtained by solving numerically
the equation $2\gamma(2/\nu, (R/a)^\nu) = \Gamma(2/\nu)$. We have
derived an analytical approximation given by:
\begin{equation}
\ln(R_{\eff}/a) = \frac{0.6950 - \ln(\nu)}{\nu} - 0.1789\, ,
\label{reff2D}
\end{equation}
which is accurate to better than 0.6\% in the range $0.1 \leq \nu \leq 1.8$
for $R_{\eff}/a$.

Another useful quantity is the Petrosian metric radius, which is related to
the logarithm slope of the brightness profile (Petrosian 1976). The
Petrosian radius is defined with the help of the variable
\begin{equation}
\eta = -2.5\log\left(\frac{\diff \log L(R)}{2\diff \log R}\right) =
\mu(R) -\langle\mu(R)\rangle\, ,
	\label{etaPetro}
\end{equation}
where $\mu(R)$ is the magnitude (equal to $-2.5\log(\Sigma)$) at $R$
and $\langle\mu(R)\rangle$ is the mean magnitude inside $R$ (Sandage \&
Perelmuter 1990).
For a given value of $\eta$, a radius $R_{p}$ is defined. Besides
being model independent, the Petrosian radius gives a metric measure
that is distance independent.
For the particular case where $R_{p} = R_{\eff}$, the Petrosian parameter
$\eta$ is a function of $\nu$ given approximately by:
\begin{equation}
\eta = 0.6213 -1.25 \log(\nu) + 0.07656 \nu\, ,
	\label{eta(nu)}
\end{equation}
which is valid in the range $0.1 \leq \nu \leq 1.8$ with an accuracy of 0.1\%.

As we mentioned before, almost all
`astrophysical' quantities such as the effective radius, the effective
brightness and the mean effective brightness are non-linear combinations
of the `primary parameters' $\Sigma_{0}$, $a$ and $\nu$.

It is also interesting to rewrite the \Se\ profile
(Eq.~\ref{sersic}) introducing explicitly the effective Radius:
\begin{equation}
\Sigma(R) = \Sigma_0 \exp\left[ - b_\nu (R/R_\eff)^\nu \right] \, ,
\label{sersic2}
\end{equation}
with $b_\nu = (R_\eff/a)^\nu$. The original \Se\ profile, Eq.~(\ref{sersic0}),
is obtained making $k_n = 1.0857 b_\nu$ and $\mu(R) = -2.5 \log\Sigma(R)$.

\subsection{Three dimensional S\'ersic deprojection}

The two dimensional \Se\ density distribution has no analytical
deprojection in three dimensions. For our purposes, numerical deprojection
(cf. Young 1976 for the de Vaucouleurs case) or the asymptotic behaviour
(Ciotti 1991) of the \Se\ profile are not well adapted. Mellier \& Mathez
(1987) have given a simple analytical approximation of the deprojection of
the de Vaucouleurs profile;
we will use a modified form of the Mellier \& Mathez three dimensional
density profile,
generalised for the \Se\ profile case:
\begin{equation}
\rho(r) = \rho_0 {\left({{r \over a}}\right)}^{-p}
\exp\left({-{\left({{r \over a}}\right)}^{\nu}}\right)\, ,
\label{3Drho}
\end{equation}
with the normalisation:
\begin{equation}
\rho_{0} = \Sigma_{0} \frac{\Gamma(2/\nu)}{2 a \Gamma((3-p)/\nu)}
\frac{M}{L}\, ,
\label{norma3D2D}
\end{equation}
which is obtained by requiring that both total masses corresponding to
Eqs.~(\ref{sersic}) and (\ref{3Drho}) be equal. $M/L$ is the mass to
luminosity ratio and we suppose hereafter that it is independent of the
galaxy radius. This is the above mentioned hypothesis (\S\ref{2.2}) that the
galaxy behaves like a one component system. Notice, however, that up to this
point we do not require that all galaxies have the same $M/L$ ratio.

The exponent $p$ in Eq.~(\ref{3Drho}) is obtained by fitting the
(numerical) Abel
deprojection of the two dimensional profile Eq.~(\ref{sersic}) to the
analytical three dimensional profile
Eq.~(\ref{3Drho}). A high quality match, better than 5\%, between the
numerical deprojection of
Eq.~(\ref{sersic}) and Eq.~(\ref{3Drho}) is obtained when:
\begin{equation}
p = 1.0 - 0.6097 \nu + 0.05463\nu^{2} \, ,
\label{p3Drho}
\end{equation}
for $0.1 \leq \nu \leq 1.8$, and the radial range $10^{-2} \leq
R/R_{\eff}\leq 10^{3}$. The discrepancy between the numerical and analytical
deprojections is more pronounced for very small and very large radii. Prugniel
\& Simien (1997) proceeded in the same fashion, using a modification of the
Mellier
\& Mathez profile, but their deprojection is only valid in the range $0.1
\le \nu \le 0.5$.
Within this range their result and ours agree well within 1\%.
For $0.5\le \nu \le 1$ our result is within 5\% to Prugniel \& Simien's, while
for $\nu > 1$, our result diverges parabolically from theirs.

Notice that the behaviour of Eq.~(\ref{3Drho})
as $r\rightarrow 0$ is not the same as the exact Abel deprojection of the
\Se\ profile. For small radii, $\rho_{Abel}(r) \propto r^{\nu-1}$ for
$\nu < 1$. This should not introduce any appreciable error in our analytical
approximation; the fraction of mass inside $10^{-2} \leq
R/R_{\eff}\leq 10^{3}$ is 97\% for $\nu=0.1$, 99.6\% for $\nu=0.25$, and
greater than 99.9\% for $\nu\ge 0.55$.

The corresponding mass inside the radius $r$, $M(r)$, is readily obtained
integrating Eq.~(\ref{3Drho}) in a spherical volume:
\begin{equation}
	M(r) = \frac{4\pi\; \rho_0 a^3}{\nu} \gamma\left(\frac{3-p}{\nu},
\left(
	\frac{r}{a}\right)^\nu\right)\, .
	\label{mass3D}
\end{equation}

Additionally, we give the relation between the two dimensional half-mass radius
$R_{\eff}$ and the three dimensional half-mass radius, $r_{\eff}$, as a
polynomial
approximation:
\begin{equation}
r_{\eff}/R_{\eff} = 1.356 - 0.0293\nu + 0.0023\nu^{2}\, ,
\label{reff3D/2D}
\end{equation}
This approximation agrees well
within 0.25\% to the values computed by Ciotti (1991) who used the
numerical Abel
inversion of the \Se\ profile. This reassures us on the quality of
Eq.~(\ref{3Drho}).

We define a gravitational radius by the relation $r_g \equiv GM^2/|U|$,
with $U$ the gravitational potential energy. Again, we give a simple
analytical approximation:
\begin{equation}
\ln(r_g/a) = 0.845 + \frac{0.820 - 0.924 \ln(\nu)}{\nu}\, ,
	\label{rg3D}
\end{equation}
that is accurate to better than 0.8\% in the range $0.1 \leq \nu \leq 1.8$
for $r_{g}/a$.
This radius is a combination of the natural scale length and the form
factor; this is naturally due to the loss of homology when moving from
a de Vaucouleurs law to a $\nu$-model.
A similar behaviour occurs with $R_\eff$,
Eq.~(\ref{reff2D}). It turns out that the relation between $R_\eff$
and $r_g$ is roughly linear, independent of $\nu$, given approximately
by
\begin{equation}
\ln(r_g/a) = 1.16 + 0.98\ln(R_\eff/a) \, ,
	\label{rg/Reff}
\end{equation}
accurate to better than 5\% in the range $0.1 \leq \nu \leq 1.8$ for $r_{g}/a$.
For  $0.15 \le \nu \le 1.0$, the relation is practically linear (within
15\%), viz.
$r_g \approx 3.0 R_\eff$ -- an interesting result given that the \Se\ model
is not homologous.

\subsection{Entropy and the S\'ersic profile}

Given a mass distribution law, the deprojected \Se\ law, we can make the
connection between the specific entropy and the parameters of the light
profile. We need, however, an additional hypothesis to get the pressure
corresponding to Eq.~(\ref{3Drho}). We consider that the pressure is
isotropic and the galaxy is in hydrostatic equilibrium. This implies an
isotropic velocity dispersion tensor and the simple relation,
$P(r) = \rho(r) \sigma_r^2$,
relating the pressure, density, and radial velocity dispersion, $\sigma_r$.
With the isotropy hypothesis, $\sigma_r$ (or $P(r)$) is, thus, obtained as a
solution of
the hydrostatic equilibrium equation:
\begin{equation}
{\diff  \over \diff r} \left[\rho(r)\;\sigma_r^2(r)\right]
 =-{G\;M(r)\;\rho(r) \over r^2}\, ,
\label{hydro}
\end{equation}

With the above hypothesis, one may substitute the density and pressure in
Eq.~(\ref{meanThermoEntro})
and a relation of the kind $s(a, \Sigma_{0}, \nu)$ can be derived. It turns
out that
$s(a, \Sigma_{0}, \nu)$ may be
well approximated by the analytical expression:
\begin{eqnarray}
s(a, \Sigma_{0}, \nu) & = &{1\over 2}\ln(\Sigma_0)+{5 \over 2}\ln(a)\nonumber\\
 & + & c_1 \ln(\nu )-\frac{1}{\nu} + c_2 \nu^{c_3} + c_0\, .
\label{Entro(a,nu,io)}
\end{eqnarray}
The constants $c_{i}$ as well as the derivation of
Eq.~(\ref{Entro(a,nu,io)}) are given in Appendix~2.

It is interesting to
note that for a Plummer profile (polytrope of index 5, e.g. Binney \& Tremaine
1987), where we are able to compute exactly the specific entropy, we have the
same dependency of $s$ on the Plummer scale and normalisation parameters as in
Eq.~(\ref{Entro(a,nu,io)}).

The relations that can be written as the formal expression
\begin{equation}
s(a, \Sigma_{0}, \nu) = s_{i}\, ,
	\label{entroplan}
\end{equation}
where $s_i$ means any given value of the specific entropy, define a family of
surfaces in the three dimensional space defined by the parameters
$\Sigma_{0}$, $a$, $\nu$. By analogy with the \FP, we call such a surface
the `Entropic Plane'.

We do not know \textit{a
priori} the value of the specific entropy. We can, nevertheless, propose two
hypotheses concerning the entropy of elliptical galaxies:

\begin{itemize}
\item \textit{Hypothesis} I: The specific entropy of early-type galaxies in a
given cluster is constant. This is the \textit{weak hypothesis}.

\item \textit{Hypothesis} II: The specific entropy of early-type
galaxies in clusters is universal, independent of the cluster in which they are
located. This is \textit{the strong hypothesis}.
\end{itemize}

One may hope that in the future, some theory of galaxy formation will
move one of the two hypotheses above into a firm knowledge and provides us
with the true value of $s$. In the absence of a way to calculate
$s$ from first principles, these hypotheses must be
tested using the observed brightness profiles of early type galaxies.

\section{The data}\label{data}

In order to verify these hypotheses, we use the brightness profile data of
elliptical galaxies belonging to two rich clusters of galaxies -- Coma and
ABCG~85 -- and the group associated with NGC~4839 (which is in the
neighbourhood of Coma).

\subsection{Description}

Galaxies in the Coma cluster have been selected from the catalogue of Biviano et 
al. (1995a) among those in the centre of the cluster and in the group around 
NGC~4839, being brighter than 17.8 $V$ mag, fainter than 13 $V$ mag (to exclude the 
biggest, saturated galaxies), with elliptical morphological types as assigned by 
Godwin, Metcalfe \& Peach (1983, hereafter GMP), and with measured redshifts 
that confirm them as cluster members (radial velocity in the range 
3000--10000~km~s$^{-1}$, cf. Biviano et al. (1995b)). The elliptical galaxies of 
ABCG~85 were extracted from the catalogue of Slezak et al. (1998), by selecting 
those galaxies belonging to the cluster (from available redshifts, with 
velocities in the range 13350--20000 km~s$^{-1}$, cf. Durret et al. (1998)) and 
with elliptical appearance in the CCD images. The final list of elliptical 
galaxies used for the fitting is given in Tables \ref{ComaSaida}, 
\ref{A85Saida}, and \ref{N4839Saida} (in their first columns we give the 
corresponding identifications in both catalogues, Biviano et al. 1995a for Coma 
and the NCG~4839 group, and Durret et al. 1998 for ABCG~85).

We have used $V$ images that had been reduced and calibrated following standard 
techniques (see Lobo et al. 1997, Slezak et al. 1998). We then cleaned the 
images by eliminating foreground stars that contaminated our target galaxies. 
This was made through specific tasks within \textsc{IDL} (Interactive Data 
Language, Research Systems, Inc.). The elimination was done by interpolating the 
field around the star by a 4th degree polynomial. We then obtained the growth 
curves (integrated magnitudes within circular regions) by means of the `ellipse' 
task in `stsdas' environment of \textsc{iraf}\footnote{\textsc{iraf} is the 
Image Analysis and Reduction Facility made available to the astronomical 
community by the National Optical Astronomy Observatories, which are operated by 
the Association of Universities for Research in Astronomy (AURA), Inc., under 
contract with the U.S. National Science Foundation.}. For every galaxy in our 
sample we obtained a table listing the radii of the circular apertures and the 
corresponding integrated magnitudes inside the apertures. Galaxies that had less 
than nine data points were rejected. This means that the galaxies on the faint 
surface brightness end were rejected by our selection. Since our analysis is not 
sensitive to the completeness of the sample, the rejection of very faint 
galaxies should have no effect in our results.

The use of a circular aperture was motivated by two reasons. First, in order to 
be able to compare directly our results with those found in the literature, in 
particular Graham \& Colless (1997) and, second, because the theory we developed 
is based on spherical symmetry of the mass distribution. Notice that previously 
(Gerbal et al. 1997) we have employed a different method for obtaining the 
growth curves, taking into account the shape of the isophotes (we computed the 
total luminosity inside a given isophote). Comparing our previous results 
(Figs.~1 and 2 from Gerbal et al. 1997) with those presented here shows that the 
difference is not significant (i.e., the pairwise distributions of the \Se\ 
parameters do not change).

\subsection{The fitting method}

We have fitted the resulting luminosity growth curves of the elliptical
galaxies in Coma (including the group of NGC~4839) and ABCG~85.

The advantage of fitting the integrated flux instead of the luminosity profile
is that the latter is more sensitive to irregularities in the light
distribution. Besides, fitting the light growth curve gives more weight to
the inner than the outer parts of the galaxy, reflecting better the galaxy
overall structure (D'Onofrio \& Prugniel 1997).
Therefore, since the inner region of a given galaxy
is probably more relaxed than the outer region (which is more susceptible to
perturbations) it seems more reasonable to give more weight to the inner
region when computing the specific entropy. The drawback is that if the galaxy
has some unusual centre, an active nucleus, star burst or some exceptional
darkening, then these features will be propagated in the whole growth curve. In
this case, fitting the luminosity profile would be better to subtract the
anomalous central contribution (D'Onofrio \& Prugniel 1997). Moreover, Burstein
et al. (1987) pointed out that the fitting of growth curve have systematic
uncertainties that may lead to correlated errors of the total magnitude and
the effective radius. This last point may be minimized by fitting the uncorrelated
primary parameters of the \Se\ law (cf. \S\ref{generalites}) rather than
their non linear combinations.

Consequently, we use the integrated form of the brightness profile
as given by Eq.~(\ref{Ltotsersic}). The luminosity growth curve fits are
done with a standard least square
minimisation method, using the `\textsc{minuit}' programme from the CERN
software library.
In order to reduce the effects of the seeing (FWHM $\approx 0.9$ arcsec for
Coma and
1.2 arcsec for ABCG~85) and/or eventual nuclear activity, the data points are
taken only from 2.5 arcsec outwards. For galaxies of the same cluster, the
effect of using a central exclusion mask fixed in angular units, rather than a
linear measure, is virtually negligible (in this precise case the galaxies
can be considered to be at the same distance from us; for Coma the error
should be smaller than $\sim5$\%).
A difficulty may rise in comparing galaxies in different clusters. However,
in our case, the clusters are not more than a factor 2.5 apart (radial
distance; cf. \S\ref{DistIndicator} below) and changing the size of the mask
from 0.6 to 3.0 arcsec did not alter significantly the fitting results,
i.e., the values obtained for the \Se\ parameters were compatible within
their 3$\sigma$ error bars [notice that this test on the size of the central
mask were also done by Gerbal et al. (1997)].

We have used the growth curve up to the surface brightness isophote at
$24.0\; V$~mag/arcsec$^2$, corresponding to a 3$\sigma$ signal above the background sky.
This is done in order to minimize errors introduced in the background
subtraction. We performed the fits varying the outer cut-off from 22.5
to 24.5 $V$~mag/arcsec$^2$; the fitting results were still basically
the same, i.e., the values obtained for the \Se\ parameters were compatible within
their 3$\sigma$ error bars.

\begin{table}
\caption{Growth curve fitting results for Coma. Ident is the GMP number
and name is either an IC or NGC identification. $\nu$ and $a$ are the shape
and
scale parameters fitted with the \Se\ law. $\mu_0=-2.5\log(\Sigma_0)$ is the
intensity parameter and the magnitude is in the $V$ band. $R_{\eff}$ and $\mumean_{\eff}$ are calculated from the
primary parameters as explained in \S\ref{generalites} }
\halign{#\hfill& \ #\hfill&\ \hfill # & \ \hfill # & \hfill # &
 \ \hfill # & \hfill #\cr
\noalign{\hrule\smallskip}
Ident$^{*}$ & name & $\nu$ & $\log(a)$ & $\mu_{0}$ &
$R_{\eff}$ & $\mumean_{\eff}$\cr
 & & & (arcsec) & (mag/$\Box\arcsec$) & (arcsec) & (mag/$\Box\arcsec$)\cr
\noalign{\smallskip\hrule\smallskip}
2727 & IC4026 & 0.27 & -2.27 & 14.33 & 6.74 & 20.58 \cr
2736 &  & 0.71 & -0.15 & 19.31 & 2.51 & 21.14 \cr
2753 &  & 0.79 & 0.00 & 19.69 & 2.73 & 21.27 \cr
2777 &  & 0.53 & -1.01 & 16.68 & 1.02 & 19.41 \cr
2794 & NGC4898B & 0.49 & -0.65 & 16.48 & 3.29 & 19.50 \cr
2798 & NGC4898A & 0.50 & -0.60 & 15.84 & 3.28 & 18.76 \cr
2805 &  & 0.32 & -2.04 & 14.27 & 2.42 & 19.46 \cr
2839 & IC4021 & 0.32 & -1.91 & 14.14 & 2.94 & 19.24 \cr
2879 &  & 0.69 & -0.15 & 19.38 & 2.73 & 21.28 \cr
2897 &  & 0.64 & -0.13 & 18.84 & 3.71 & 20.97 \cr
2910 &  & 1.09 & 0.23 & 18.82 & 2.48 & 19.79 \cr
2910 &  & 0.50 & -0.62 & 16.67 & 3.23 & 19.61 \cr
2914 &  & 0.43 & -0.67 & 18.35 & 6.42 & 21.92 \cr
2921 & NGC4889 & 0.45 & -0.32 & 16.10 & 10.86 & 19.45 \cr
2922 & IC4012 & 0.53 & -0.66 & 15.96 & 2.30 & 18.71 \cr
2940 & IC4011 & 0.34 & -1.53 & 15.37 & 4.05 & 20.07 \cr
2960 &  & 0.43 & -0.80 & 17.53 & 4.62 & 21.09 \cr
2975 & NGC4886 & 0.30 & -1.97 & 14.35 & 5.48 & 19.99 \cr
2976 &  & 0.63 & -0.34 & 19.10 & 2.34 & 21.25 \cr
3058 &  & 1.07 & 0.34 & 20.46 & 3.31 & 21.48 \cr
3073 & NGC4883 & 0.44 & -0.81 & 16.10 & 4.19 & 19.59 \cr
3084 &  & 0.42 & -1.00 & 16.27 & 3.24 & 19.90 \cr
3113 &  & 0.66 & -0.21 & 19.15 & 2.73 & 21.18 \cr
3126 &  & 0.87 & 0.10 & 19.56 & 2.73 & 20.93 \cr
3133 &  & 0.46 & -0.91 & 17.19 & 2.64 & 20.50 \cr
3170 & IC3998 & 0.33 & -1.52 & 15.52 & 5.55 & 20.43 \cr
3170 & IC3998 & 0.34 & -1.47 & 15.50 & 4.89 & 20.24 \cr
3201 & NGC4876 & 0.52 & -0.53 & 16.65 & 3.36 & 19.46 \cr
3206 &  & 0.54 & -0.47 & 17.56 & 3.19 & 20.21 \cr
3213 &  & 0.42 & -1.07 & 15.95 & 2.98 & 19.64 \cr
3222 &  & 0.48 & -1.00 & 15.73 & 1.61 & 18.82 \cr
3222 &  & 0.93 & -0.08 & 17.61 & 1.59 & 18.85 \cr
3269 &  & 0.34 & -1.83 & 14.87 & 2.50 & 19.73 \cr
3291 &  & 0.33 & -1.42 & 16.92 & 7.26 & 21.86 \cr
3292 &  & 0.64 & -0.23 & 19.20 & 2.97 & 21.34 \cr
3296 & NGC4875 & 0.39 & -1.29 & 15.17 & 2.92 & 19.22 \cr
3302 &  & 0.45 & -0.93 & 17.32 & 2.66 & 20.66 \cr
3312 &  & 0.51 & -0.89 & 18.11 & 1.60 & 20.99 \cr
3329 & NGC4874 & 0.60 & 0.20 & 17.70 & 10.01 & 20.01 \cr
3340 &  & 1.07 & 0.13 & 20.77 & 2.04 & 21.77 \cr
3352 & NGC4872 & 0.50 & -0.72 & 15.77 & 2.56 & 18.71 \cr
3367 & NGC4873 & 0.35 & -1.40 & 15.40 & 5.19 & 20.07 \cr
3367 & NGC4873 & 0.30 & -1.82 & 14.80 & 6.67 & 20.32 \cr
3400 & IC3973 & 0.34 & -1.82 & 13.65 & 2.55 & 18.51 \cr
3414 & NGC4871 & 0.23 & -3.17 & 12.63 & 6.12 & 20.19 \cr
3423 & IC3976 & 0.32 & -2.01 & 13.61 & 2.59 & 18.78 \cr
3439 &  & 0.62 & -0.19 & 18.60 & 3.56 & 20.81 \cr
3486 &  & 0.78 & -0.26 & 18.79 & 1.55 & 20.42 \cr
3487 &  & 0.41 & -1.27 & 15.94 & 2.29 & 19.78 \cr
3489 &  & 0.91 & -0.04 & 19.99 & 1.83 & 21.29 \cr
3510 & NGC4869 & 0.47 & -0.66 & 16.02 & 4.24 & 19.25 \cr
3522 &  & 0.43 & -1.14 & 15.61 & 2.14 & 19.17 \cr
3534 &  & 0.58 & -0.43 & 18.23 & 2.71 & 20.67 \cr
3554 &  & 0.37 & -1.33 & 16.80 & 3.99 & 21.16 \cr
3557 &  & 0.51 & -0.48 & 17.54 & 3.98 & 20.40 \cr
3561 & NGC4865 & 0.43 & -0.89 & 15.15 & 3.70 & 18.69 \cr
3565 &  & 0.65 & -0.23 & 19.85 & 2.84 & 21.94 \cr
3639 & NGC4867 & 0.54 & -0.57 & 16.03 & 2.59 & 18.70 \cr
\noalign{\smallskip\hrule\smallskip} }
${}^*$ From the catalogue of Biviano et al. (1995a)
\label{ComaSaida}
\end{table}

\begin{table}
\noindent{\bf Table \ref{ComaSaida}.} Continued \\
\halign{#\hfill& \ #\hfill&\ \hfill # & \ \hfill # & \hfill # &
 \ \hfill # & \hfill #\cr
\noalign{\hrule\smallskip}
Ident$^{*}$ & name & $\nu$ & $\log(a)$ & $\mu_{0}$ &
$R_{\eff}$ &  $\mumean_{\eff}$\cr
 & & & (arcsec) & (mag/$\Box\arcsec$) & (arcsec) & (mag/$\Box\arcsec$)\cr
\noalign{\smallskip\hrule\smallskip}
3656 &  & 0.20 & -3.55 & 13.65 & 16.07 & 22.39 \cr
3664 & NGC4864 & 0.49 & -0.55 & 16.25 & 4.26 & 19.28 \cr
3681 &  & 0.62 & -0.34 & 18.92 & 2.53 & 21.13 \cr
3707 &  & 0.47 & -0.74 & 18.02 & 3.31 & 21.19 \cr
3733 & IC3960 & 0.25 & -2.79 & 13.27 & 5.67 & 20.21 \cr
3782 &  & 0.33 & -1.79 & 15.05 & 3.38 & 20.06 \cr
3792 & NGC4860 & 0.45 & -0.74 & 15.58 & 4.03 & 18.91 \cr
3794 &  & 0.50 & -0.87 & 16.59 & 1.89 & 19.56 \cr
3855 &  & 0.63 & -0.20 & 19.54 & 3.36 & 21.71 \cr
3914 &  & 0.56 & -0.70 & 16.11 & 1.68 & 18.67 \cr
4103 &  & 0.74 & -0.02 & 19.43 & 3.07 & 21.16 \cr
4129 &  & 1.04 & 0.07 & 19.88 & 1.86 & 20.94 \cr
4200 &  & 0.55 & -0.61 & 16.87 & 2.15 & 19.47 \cr
4230 &  & 0.23 & -3.05 & 12.81 & 7.90 & 20.36 \cr
\noalign{\smallskip\hrule\smallskip} }
${}^*$ From the catalogue of Biviano et al. (1995a)
\end{table}

\begin{table}
\caption{Growth curve fitting results for ABCG~85 The columns have the
same meaning as in Table~\ref{ComaSaida}, except for the Ident column.}
\halign{#\hfill & \hfill # & \quad \hfill # & \hfill # &
 \ \hfill # & \hfill #\cr
\noalign{\hrule\smallskip}
Ident$^{*}$ & $\nu$ & $\log(a)$ & $\mu_{0}$ &
$R_{\eff}$ &  $\mumean_{\eff}$\cr
 & & (arcsec) & (mag/$\Box\arcsec$) & (arcsec) & (mag/$\Box\arcsec$)\cr
\noalign{\smallskip\hrule\smallskip}
152 & 0.59 & -0.63 & 17.9 & 1.56 & 20.30 \cr
156 & 0.20 & -4.40 & 13.0 & 3.52 & 22.00 \cr
175 & 0.29 & -2.49 & 14.7 & 2.14 & 20.56 \cr
179 & 0.50 & -0.92 & 17.7 & 1.58 & 20.68 \cr
182 & 0.69 & -0.02 & 18.6 & 3.69 & 20.50 \cr
197 & 0.38 & -1.31 & 15.8 & 3.05 & 19.95 \cr
202 & 0.71 & -0.30 & 18.2 & 1.81 & 20.05 \cr
208 & 0.26 & -2.61 & 15.5 & 5.91 & 22.20 \cr
209 & 0.76 & -0.07 & 18.9 & 2.53 & 20.60 \cr
212 & 0.64 & -0.58 & 18.3 & 1.30 & 20.42 \cr
214 & 0.46 & -0.62 & 18.1 & 4.94 & 21.40 \cr
218 & 1.16 & 0.12 & 20.6 & 1.79 & 21.52 \cr
222 & 0.63 & -0.54 & 18.5 & 1.51 & 20.67 \cr
225 & 0.71 & -0.32 & 18.7 & 1.74 & 20.54 \cr
228 & 0.89 & -0.07 & 20.2 & 1.78 & 21.48 \cr
229 & 0.80 & -0.21 & 20.1 & 1.66 & 21.71 \cr
231 & 0.46 & -1.03 & 20.5 & 1.82 & 23.78 \cr
235 & 0.49 & -0.87 & 17.7 & 1.93 & 20.74 \cr
236 & 0.44 & -0.96 & 16.3 & 2.70 & 19.77 \cr
246 & 1.43 & 0.31 & 22.8 & 2.18 & 23.40 \cr
253 & 0.62 & -0.48 & 18.3 & 1.77 & 20.49 \cr
263 & 0.24 & -3.59 & 13.1 & 1.49 & 20.36 \cr
283 & 0.34 & -1.49 & 16.8 & 5.35 & 21.67 \cr
296 & 0.85 & 0.01 & 19.6 & 2.30 & 20.98 \cr
305 & 0.76 & -0.19 & 19.4 & 1.93 & 21.05 \cr
310 & 0.94 & -0.12 & 19.8 & 1.43 & 21.04 \cr
316 & 1.05 & 0.25 & 19.9 & 2.77 & 20.92 \cr
324 & 0.35 & -1.31 & 17.9 & 6.27 & 22.54 \cr
329 & 0.66 & -0.42 & 19.4 & 1.70 & 21.39 \cr
\noalign{\smallskip\hrule\smallskip} }
${}^*$ From the catalogue of Durret et al. (1998)
\label{A85Saida}
\end{table}

\begin{table}
\caption{Growth curve fitting results for the N4839 group. The columns have
the
same meaning as in Table~\ref{ComaSaida}}
\halign{#\hfill& \ #\hfill&\ \hfill # & \ \hfill # & \hfill # &
 \ \hfill # & \hfill #\cr
\noalign{\hrule\smallskip}
Ident$^{*}$ & name & $\nu$ & $\log(a)$ & $\mu_{0}$ &
$R_{\eff}$ &  $\mumean_{\eff}$\cr
 & & & (arcsec) & (mag/$\Box\arcsec$) & (arcsec) & (mag/$\Box\arcsec$)\cr
\noalign{\smallskip\hrule\smallskip}
4714 &  & 0.48 & -0.83 & 17.29 & 2.38 & 20.38 \cr
4792 & NGC4842B & 0.52 & -0.62 & 16.65 & 2.62 & 19.42 \cr
4794 & NGC4842A & 0.49 & -0.61 & 16.05 & 3.75 & 19.10 \cr
4852 &  & 0.69 & -0.32 & 18.60 & 1.84 & 20.49 \cr
4907 &  & 0.41 & -1.17 & 15.26 & 2.75 & 19.06 \cr
4928 & NGC4839 & 0.25 & -2.03 & 14.63 & 27.87 & 21.46 \cr
4937 &  & 1.07 & 0.26 & 20.60 & 2.78 & 21.62 \cr
5051 &  & 0.32 & -1.77 & 14.17 & 3.89 & 19.24 \cr
5102 &  & 0.57 & -0.22 & 19.07 & 4.65 & 21.56 \cr
\noalign{\smallskip\hrule\smallskip} }
${}^*$ From the catalogue of Biviano et al. (1995a)
\label{N4839Saida}
\end{table}

\section{Fitting results and correlations}\label{Correlations}

\begin{figure}
\centerline{\psfig{figure=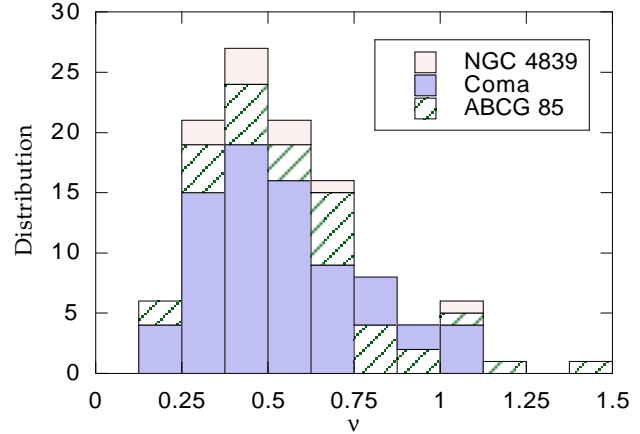,width=8.4truecm}}
\caption{Distribution of the shape parameter $\nu$ for all galaxies,
grouped by cluster membership. The
distribution does not peak at 0.25, the value corresponding to the de
Vaucouleurs profile.}
\protect\label{figNuDist}
\end{figure}

\begin{figure}
\centerline{\psfig{figure=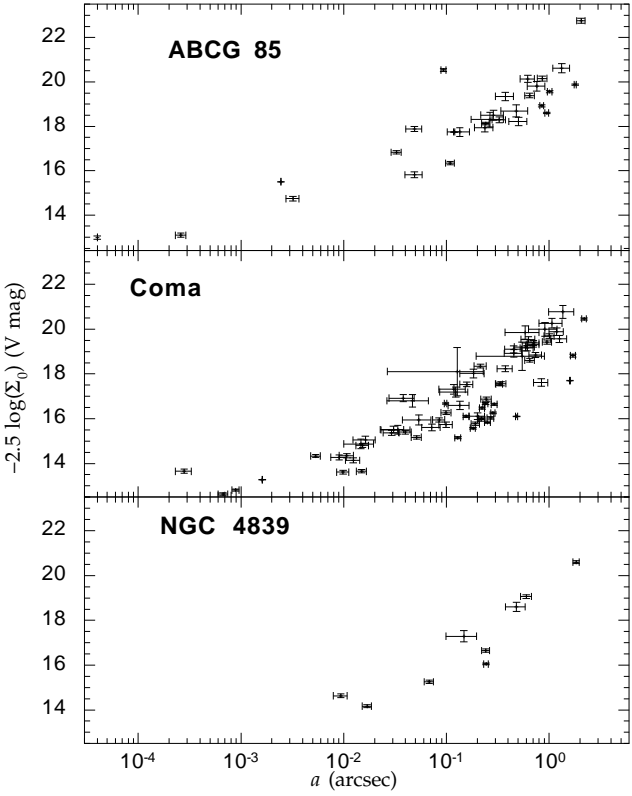,width=8.4truecm}}
\caption{Distribution of the intensity parameter $-2.5 \log \Sigma_0$
(in $V$ magnitude) and the scale length $a$ (in
arcsec) for all galaxies, each panel for a cluster. The bars are
$1\sigma$ errors.}
\protect\label{figSig0a}
\end{figure}

\begin{figure}
\centerline{\psfig{figure=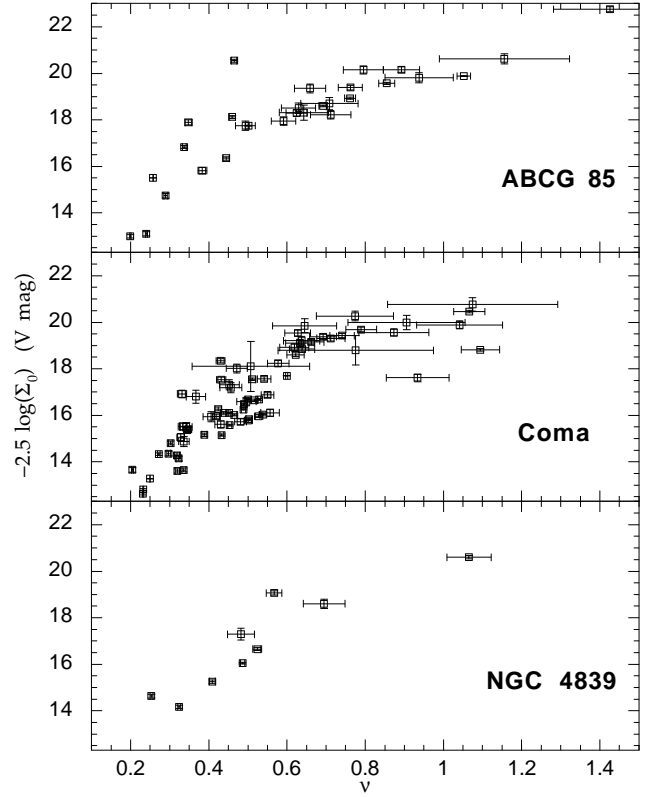,width=8.4truecm}}
\caption{Distribution of $-2.5 \log \Sigma_0$(in $V$ magnitude) and $\nu$.
The bars are $1\sigma$ errors.}
\protect\label{figSig0Nu}
\end{figure}

\begin{figure}
\centerline{\psfig{figure=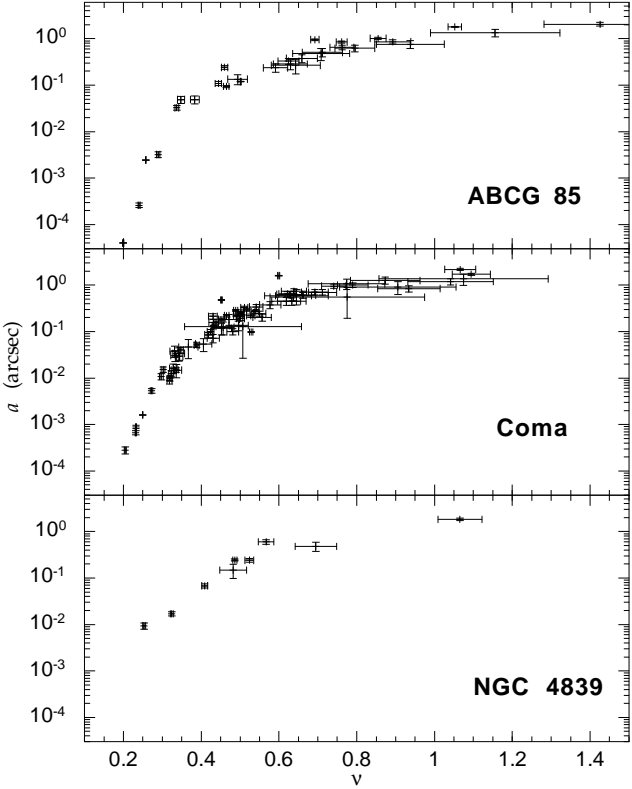,width=8.4truecm}}
\caption{Distribution of $a$ (in arcsec) and $\nu$. Notice that here we
have the
best correlation between two of the three parameters of the \Se\ profile.
The bars are $1\sigma$ errors.}
\protect\label{figANu}
\end{figure}

\begin{figure}
\centerline{\psfig{figure=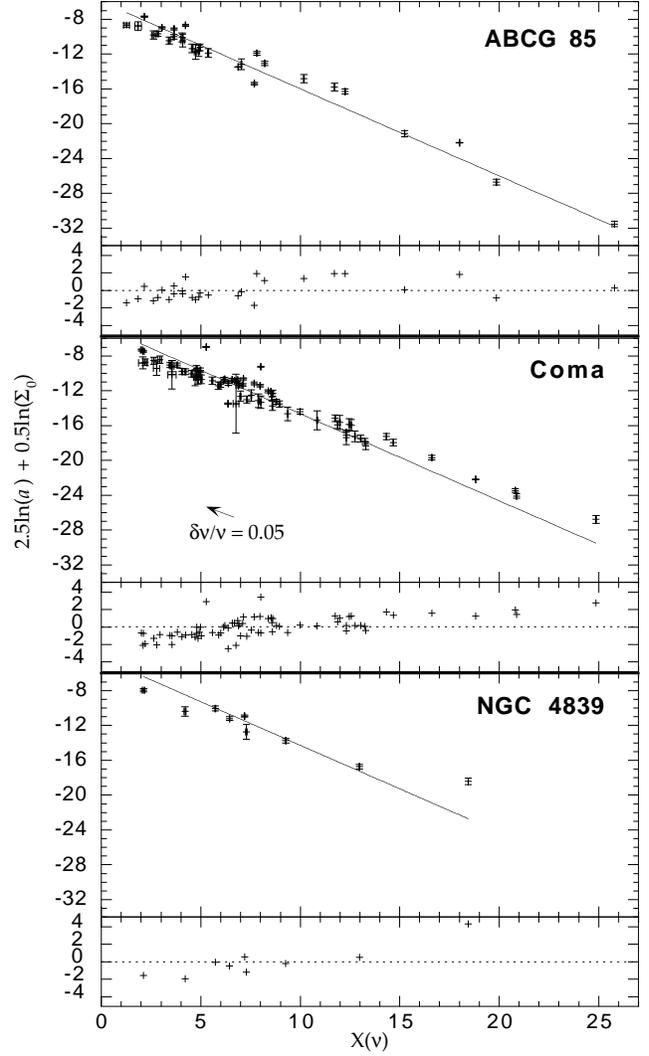,width=8.4truecm}}
\caption[]{The specific entropy relation and their corresponding residual for the
elliptical galaxies in Coma, ABCG~85, and the group of NGC~4839. The straight
lines are the `Entropic Lines', (Eq.~(\ref{EntroLin}), where the slope is equal
to $-1$ and the zero-points, $s_0$, are from Table~\ref{stat}: $-5.99\pm0.20$,
$-4.63\pm0.14$, and $-4.28\pm0.61$ for ABCG~85, Coma and NGC~4893, respectively.
Note that $X(\nu)$ is a monotonic descending function of $\nu$, i.e., high
values of $X(\nu)$ correspond to small $\nu$. The residual plot below each Entropic 
Line is $Y_{\rm data} - Y_{\rm predicted}$. The arrow in the Coma panel shows 
the typical displacement vector for an error in $\nu$ of 5\%. This displacement 
is due to the definition of $X(\nu)$ and the correlations with $\Sigma_{0}$ and $a$.}
\protect\label{figEntroLin}
\end{figure}

\begin{figure}
\centerline{\psfig{figure=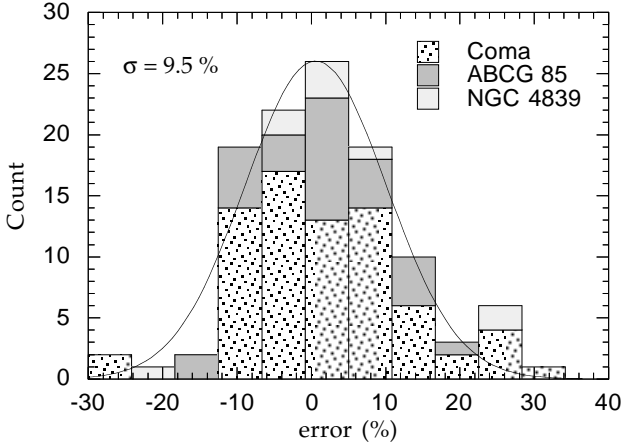,width=8.4truecm}}
\caption[]{Dispersion distribution of the galaxies around the `Entropy
Line' shown
in Fig.~\ref{figEntroLin}. This error is associated with the uncertainty in
determining the specific entropy of each galaxy (see \S\ref{Correlations}) and
the relative distances between clusters (see \S\ref{DistIndicator}). We show
also the best fit (standard least-square method) gaussian to all data points
and the one standard error, $\sigma$, in percent.}
\protect\label{erroSdist}
\end{figure}

\begin{figure}
\centerline{\psfig{figure=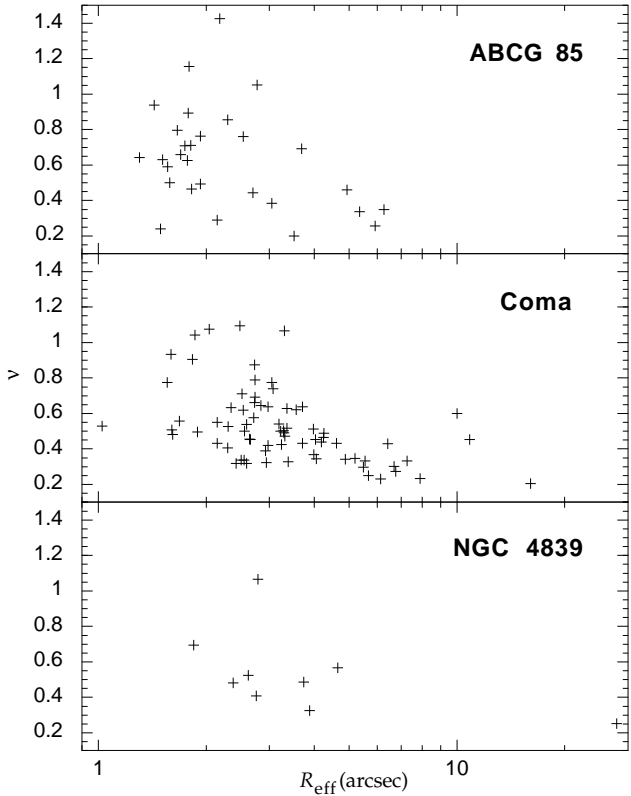,width=8.4truecm}}
\caption[]{Scatter diagram of $\nu$ and $R_\eff$. Notice the weaker
correlation in
these plots compared to the correlations between the primary parameters
$\nu$ and $a$.}
\protect\label{figReffnu}
\end{figure}

\begin{figure}
\centerline{\psfig{figure=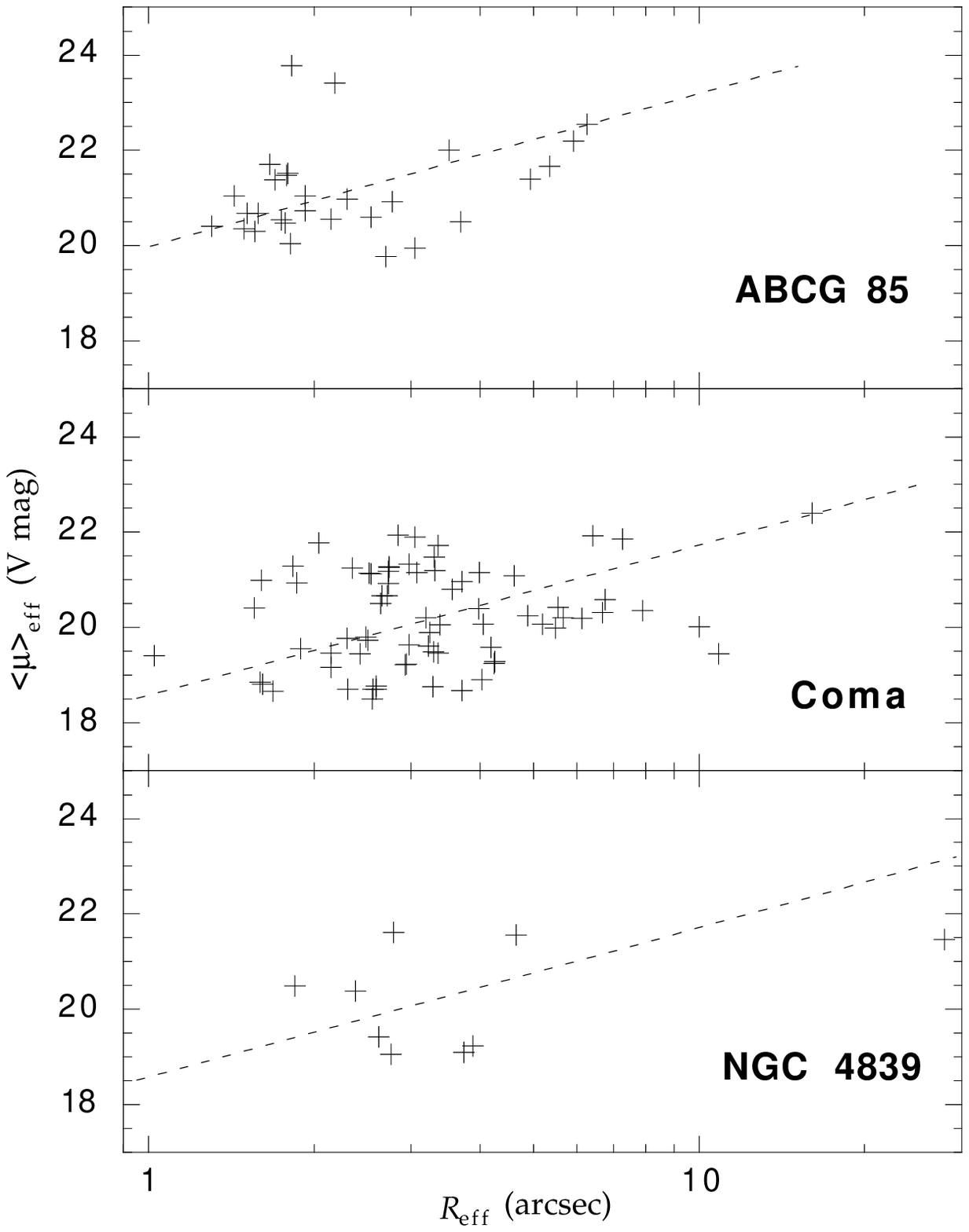,width=8.4truecm}}
\caption[]{Scatter diagram of $\mumean_{\eff}$ and $R_\eff$. The dotted
lines are
from the relation $\log R_\eff \propto 0.32 \mumean_{\eff}$ found by Oegerle \&
Hoessel (1991) for brightest cluster members. Notice the weaker correlation in
these plots compared to the correlations between the primary parameters
$\Sigma_0$ and $a$.}
\protect\label{figReffMu}
\end{figure}

The results of the growth curve fits are shown in Tables \ref{ComaSaida},
\ref{A85Saida}, and \ref{N4839Saida} for Coma, ABCG~85, and the group of
NGC~4839, respectively. We present also in these tables the usually quoted
effective radius, $R_\eff$, and mean surface brightness $\mumean_{\eff}$, both
computed with formulas (\ref{reff2D}) and (\ref{mu_eff}), respectively. The
errors in the \Se\ parameters are represented graphically in the Figures
\ref{figSig0a}--\ref{figANu}.

In Fig.~\ref{figNuDist} we show the distribution of the shape parameter $\nu$
for all our fits. The distribution is fairly regular and unimodal, but it
is not
symmetric (skewness $\sim 1$). It is interesting to note that $\nu=0.25$ --
the
value corresponding to the de Vaucouleurs profile -- is neither the median
($\nu=0.50$) nor the mean ($\nu=0.55$) of the distribution. The maximum of the
distribution corresponds to $\nu \sim 0.4$. Very high values of $\nu$ are
obtained for a few galaxies. Although the galaxies selected by us are
classified
as ellipticals, our fits suggest that some of them may be in fact either
lenticulars or dwarf spheroidal galaxies (those with $\nu \ga 1$). Notice,
however, that our sample is not complete (we had to reject a few faint
galaxies, the brightest saturated ones, 
and we selected only the ones with known redshift) and the distribution
will be
underestimated for the low surface brightness galaxies (i.e. high $\nu$).

In Figs.~\ref{figSig0a}, \ref{figSig0Nu}, and \ref{figANu} we display the
relations between the primary parameters $\Sigma_0$, $a$, and $\nu$. It is
obvious that all these quantities are strongly pairwise correlated, although
non-linearly. Before analysing quantitatively these correlations, let's make
the connection with the Entropic Plane.

From Eq.~(\ref{Entro(a,nu,io)}), we expect that there are indeed non-linear
correlations among the \Se\ profile parameters, since that equation defines a
surface in a three dimensional space. In fact non-linear correlations are
expected
because the Entropic Plane relates non-linearly these parameters (if the \textit{weak} or \textit{strong} hypotheses hold). In order to verify whether or not
Eq.~(\ref{Entro(a,nu,io)}) actually holds, we have defined two auxiliary
variables:
\begin{equation}
Y(\Sigma_{0},a) \equiv \frac{1}{2}\ln(\Sigma_{0}) + \frac{5}{2}\ln(a)\, ,
	\label{Y}
\end{equation}
and
\begin{equation}
X(\nu) \equiv c_{0} + c_{1}\ln(\nu) - \frac{1}{\nu} + c_{2}\nu^{c_{3}}\, ,
\label{X}
\end{equation}
(the constants $c_i$ are defined in Appendix 2). Notice that $X(\nu)$ is a
monotonic descending function of $\nu$ only.
With the above definitions, we can rewrite the specific entropy as a
simple linear relation as:
\begin{equation}
Y(\Sigma_{0},a) = s_{0} - X(\nu) \, .
\label{EntroLin}
\end{equation}

This relation defines a family of straight lines -- `Entropic Lines' which
correspond to the Entropic Plane seen edge-on -- for which the slope is $-1$
and the zero-point value (i.e. $s_{0}$, the specific entropy) is unknown.

It is straightforward to calculate $Y_{i}$ and $X_{i}$ from the values of 
$\Sigma_0$, $a$, and $\nu$ for the galaxy $i$ and, consequently, the specific 
entropy, $s_{0}$ for each galaxy (but notice that $s_{0}$ calculated in this way 
is distance dependent -- we will develop this in \S\ref{DistIndicator}). In 
Table \ref{stat}, we give for each cluster the mean value of $s_{0}$, as well as 
the standard error and deviation. The zero-point of the Entropic Lines is simply 
the mean value of $s_{0}$ shown in Table \ref{stat}.

In Fig.~\ref{figEntroLin} we display $Y_{i}$ and $X_{i}$ for all our
galaxies; the Entropic Line is superimposed to the data points, where we
have used the values quoted in Table \ref{stat}. We also show the displacement 
vector for a typical error in $\nu$ ($\delta\nu/\nu=0.05$); this is done taken 
in account the correlations of $\nu$ with both $\Sigma_{0}$ and $a$ (cf. 
Figs~\ref{figSig0Nu} and \ref{figANu}, as well de definition of $X(\nu)$).

\begin{table}
\caption[]{Statistic of the specific entropy. The mean value of the specific
entropy, $s_{0}$, is used as the zero-point of the Entropic
Line (cf. Fig.~\ref{figEntroLin}). While the Standard Deviation reflects the
individual scatter, the Standard Error takes into account the number of data
points.}
\begin{tabular}{l r r r}
\hline
	 & Coma & Abell85 & NGC4839\\
\hline
Mean $s_0$ & -4.63 & -5.99 & -4.28\\
Standard Error & 0.14 & 0.20 & 0.61\\
Standard Deviation & 1.20 & 1.09 & 1.84\\
Number of galaxies & 73 & 29 & 9\\
\hline
\end{tabular}
\label{stat}
\end{table}

The galaxies follow remarkably the \textit{predicted} linear correlation, in 
spite of all our simplifying hypotheses. The small dispersion around the line 
representing Eq.~(\ref{EntroLin}) strongly suggests that in each individual 
cluster, these galaxies share the same specific entropy: at least the \textit{weak 
hypothesis} is fulfilled. In Fig.~\ref{erroSdist}, we show the residual 
distribution, i.e., $(Y_{i} - Y_{i,\rm theor})/Y_{i,\rm theor}$ (where $Y_i$ is 
obtained from the observational data and we use Eq.~(\ref{EntroLin}) to obtain 
$Y_{i,\rm theor}$ from $s_{0} - X_{i}$); the standard deviation is equal to 
9.5\%. Applying the Kolmogorov-Smirnov test to the residual distribution shown 
in Fig.~\ref{erroSdist} results that its distribution is compatible to a 
gaussian one. Notice, however, that due to the small number of data points, 
only a very deviant distribution would not be compatible with a gaussian.

It would be possible to reduce the scatter of $X_{i}$ and $Y_{i}$ by using a 
linear fit (slope and zero-point as free parameters) or a polynomial one instead 
of computing $s_{0}$ the way we did and using the predicted Entropic Line. This, 
however, would bury the physics behind the Entropic Plane relation -- 
in fact, the variables $X(\nu)$ and $Y(\Sigma_{0}, a)$ would lose 
their meaning.
Anyway, fitting the $X_{i}$ and $Y_{i}$ data points to a free-slope straight line
(using a standard least-square method) yields the following slopes:
$-0.84\pm0.05$, $-0.93\pm0.06$, and $-0.66\pm0.12$ for Coma, Abell~85, and
NGC~4839 group, respectively. The mean value of the slope, taking into account
the error bars, is $0.86\pm0.14$. Although this is statistically consistent with
the predicted Entropic Line, the tilt is almost certainly real. There is a
number of effects that may account for the scatter and the deviation of the data
points from a line of slope exactly equal to $-1$. This point will be addressed in
\S\ref{discussion}.

\begin{table}
	\caption[]{Non parametric correlation tests between the parameters of the
	\Se-law for the Coma cluster. Both $\rho$ and $\tau$ are defined in the
	interval $[-1,1]$, the value 0 meaning no correlation. Higher absolute
	values of $z$ indicate a greater significance of the correlations.}
	\begin{tabular}{l r r r r}
\hline
	Parameters & $|\rho|$ Spearman & $|z|$ & $|\tau|$ Kendall & $|z|$ \\
\hline
	$X(\nu)$--$Y(\Sigma_{0},a)$ & 0.94 & 7.97 & 0.82 & 10.3  \\
	$\nu$--$a$ & 0.95 & 8.10 & 0.84 & 10.5  \\
	$\nu$--$\Sigma_0$ & 0.89 & 7.57 & 0.73 & 9.18  \\
	$a$--$\Sigma_0$ & 0.89 & 7.56 & 0.73 & 9.18  \\
	$\nu$--$R_{\eff}$ & 0.51 & 4.36 & 0.37 & 4.65 \\
	$R_{\eff}$--$\mumean_{\eff}$ & 0.18 & 1.54 & 0.13 & 1.60 \\
\hline
\end{tabular}
\label{nonParamComa}
\end{table}

\begin{table}
	\caption{Same as Table \ref{nonParamComa} for ABCG 85.}
	\begin{tabular}{l r r r r}
\hline
	Parameters & $|\rho|$ Spearman & $|z|$ & $|\tau|$ Kendall & $|z|$ \\
\hline
	$X(\nu)$--$Y(\Sigma_{0},a)$ & 0.95 & 5.03 & 0.84 & 6.38 \\
	$\nu$--$a$ & 0.97 & 5.13 & 0.88 & 6.68  \\
	$\nu$--$\Sigma_0$ & 0.89 & 4.71 & 0.80 & 6.04  \\
	$a$--$\Sigma_0$ & 0.85 & 4.52 & 0.74 & 5.63  \\
	$\nu$--$R_{\eff}$ & 0.32 & 1.67 & 0.22 & 1.65 \\
	$R_{\eff}$--$\mumean_{\eff}$ & 0.31 & 1.64 & 0.21 & 1.61 \\
\hline
\end{tabular}
\label{nonParamA85}
\end{table}

\begin{table}
	\caption{Same as Table \ref{nonParamComa} for NGC 4839 group.}
	\begin{tabular}{l r r r r}
\hline
	Parameters & $|\rho|$ Spearman & $|z|$ & $|\tau|$ Kendall & $|z|$ \\
\hline
	$X(\nu)$--$Y(\Sigma_{0},a)$ & 0.97 & 2.73 & 0.89 & 3.34 \\
	$\nu$--$a$ & 0.97 & 2.73 & 0.89 & 3.34 \\
	$\nu$--$\Sigma_0$ & 0.92 & 2.59 & 0.78 & 2.92 \\
	$a$--$\Sigma_0$ & 0.92 & 2.59 & 0.78 & 2.92 \\
	$\nu$--$R_{\eff}$ & 0.42 & 1.18 & 0.33 & 1.25 \\
	$R_{\eff}$--$\mumean_{\eff}$ & 0.20 & 0.56 & 0.06 & 0.21 \\
\hline
\end{tabular}
\label{nonParamSous}
\end{table}

Similar to our Fig.~\ref{figANu}, Young \& Currie (1995) have already presented 
the $\nu$--$a$ correlation for dwarf ellipticals in the direction of the Fornax 
cluster (figure 1 in their paper) and fitted it with a polynomial having five 
free parameters. Their small r.m.s. scatter (0.108) suggested that the 
$\nu$--$a$ correlation could be a useful distance indicator. More recently, 
Binggeli \& Jerjen (1998), who have also fitted the growth curve using the \Se\ 
profile, have presented the same correlation between the shape parameter and the 
scale length for early-type galaxies belonging to the Virgo cluster (Fig.~7 in 
their paper). They have fitted a parabola to the $\nu$--$a$ correlation and have 
derived a r.m.s. equal to 17\%.

We have mentioned in Section~\ref{generalites} that the `astrophysical' 
parameters are non-linear combinations of the primary parameters. It is 
therefore not surprising that the correlations between these quantities are less 
clear compared to the corresponding relations obtained when one uses the primary 
parameters. For instance, the distribution of $R_\eff$ and $\nu$ 
(Fig.~\ref{figReffnu}) compared to the distribution of $a$ and $\nu$ 
(Fig.~\ref{figANu}) shows how the primary parameters correlates better than the 
classical astrophysical parameters. One can also see the Fig.~3 from the paper 
by Graham et al. (1996), where they plot the $R_\eff$--$\nu$ relation and 
compare it with the $a$--$\nu$ relation displayed in Fig.~7 from the paper of 
Binggeli \& Jerjen (1998) for Virgo galaxies. Finally, one can compare the 
computed values of $R_\eff$ and $\mumean_{\eff}$ (Fig.~\ref{figReffMu}) and 
compare it to the relation between $-2.5\log\Sigma_{0}$ and $a$ 
(Fig.~\ref{figSig0Nu}).

As already noted by Oegerle \& Hoessel (1991), the scatter is rather larger for 
normal elliptical galaxies than for the brightest cluster members (BCG). In 
Fig.~\ref{figReffMu} we have superposed the empirical relation $\log R_\eff 
\propto 0.32 \mumean_{\eff}$ found by Oegerle \& Hoessel (1991, their Fig.~5) 
for their sample of BCGs. It is clear from Fig.~\ref{figReffMu} that this is the 
most scattered relation.

The relations observed in Figs.~\ref{figSig0a}, \ref{figSig0Nu}, and 
\ref{figANu} are nothing more than projections of the Entropic Plane predicted 
by Eq.~(\ref{Entro(a,nu,io)}) onto the three planes defined by the parameters 
$\Sigma_0$, $a$, and $\nu$. The thickness of the pairwise correlations seen in 
these figures is largely due to the fact that the specific Entropic Plane is not 
seen exactly edge-on. The non-linear combinations of these parameters (the 
$R_\eff$--$\mumean_\eff$ correlation, for instance) present, as one would 
expect, greater scatter. By the same token, one could na\"{\i}vely expect that 
the scatter of the $X_{i}$--$Y_{i}$ relation (which is also a non-linear 
combination of the \Se\ parameters) would be greater than the scatter of the 
primary parameters pairwise correlations.

In Tables \ref{nonParamComa} to \ref{nonParamSous}, we quantify the quality of 
the \Se\ parameters pairwise relations using two non-parametric analysis: the 
Spearman and the Kendall rank correlation tests (for a detailed description, see 
Siegel \& Castellan 1989; Press et al. 1992). (In \S\ref{DistIndicator} below, we 
will give the scatter around the Entropic Line in terms of the distance modulus 
for each galaxy in Fig.~\ref{fig:mu0}.) As it can be seen, the relations involving 
the secondary parameters, $R_\eff$ and $\mumean_\eff$, have indeed smaller correlation 
coefficients than the relations involving only the primary \Se\ parameters.

The Entropic Line turns out to have about the same correlation 
coefficients as the $\nu$--$a$ relation (the latter being slightly more tightly 
correlated). This tells us that the Entropic Line does not seem to be just an 
arbitrary non-linear combination of the primary parameters, but may be indeed a 
fundamental relation based on the assumption of unique specific entropy. It 
also tells us that the $\nu$--$a$ relation is almost an edge-on projection of the 
Entropic Plane. The fact that the Entropic Line has not a better correlation 
coefficient may be due to one or some of our simplifying hypotheses in deriving 
the specific entropy relation for elliptical galaxies (for instance, the ideal 
gas, the isotropic pressure, and the mass--luminosity ratio, see also 
\S\ref{discussion}).

\section{Towards a distance indicator}\label{DistIndicator}

\subsection{Distance dependence}

The specific entropy can be written as a function of the three
parameters of the \Se\ profile (cf. Eq.~\ref{Entro(a,nu,io)}). If we could
somehow
determine the value of the specific entropy for a given galaxy, then
we would have a relation linking one of the parameters as a function
of the remaining two others, for instance, $\nu(a, \Sigma_{0})$.

Any relation that operates between a dimensionless parameter and one or more
distance-dependent parameters can be potentially used as a distance
indicator. In the present case, a correlation between the distance
independent $\nu$ and the distance dependent $a$ and $\Sigma_{0}$
could be used as a yardstick.

However, a difficulty arises since we need to know \textit{a priori} the
value of the specific entropy. This problem can be solved partially if we
consider relative distances between clusters of galaxies (but not
individual galaxies) instead of absolute distances. Measuring the
distances of clusters instead of individual galaxies reduces the error
induced by the dispersion in the specific entropy relation.

The parameters $a$ and $\Sigma_{0}$ are, so far, given in metric
units, not angular. The distance dependence of $Y$, defined in
Eq.~(\ref{Y}), can be written explicitly by converting $a$ and $\Sigma_{0}$
to those observed, $a\arcsec$ and $\Sigma_{0}\arcsec$, as follows:
\begin{equation}
Y = \frac{1}{2}\ln(\Sigma_{0}\arcsec) + \frac{5}{2}\ln(a\arcsec)
+ \frac{3}{2}\ln(D)\,
\label{YDist}
\end{equation}
where $D$ is the distance to the cluster.

The fits of the growth curves provide us with the parameters 
$\Sigma_{0}\arcsec$, $a\arcsec$, and $\nu$, thus the parameters $X$ and $Y$ 
(without the distance term). Given a sample of early type galaxies from a 
cluster of known distance, we obtain the specific entropy as the mean value of 
$s_0$ found for all sample galaxies in that cluster, provided that Hypothesis I 
holds.

\subsection{Relative distances}

Likewise, given two samples of galaxies on two different clusters
of unknown distances, and assuming that Hypothesis II holds, we could
determine the relative distance of the clusters directly from 
Eq.~(\ref{YDist}).
The relative distance between two clusters is
given by
\begin{equation}
\frac{D_{1}}{D_{2}} = \exp\left(\frac{s_{2} - s_{1}}{1.5}\right) \, ,
\label{distRelative}
\end{equation}
where $s_{i}$'s are the mean entropy of each cluster. To these distance
corresponds the relative distance modulus, $\mu$:
\begin{equation}
\mu _{1}-\mu _{2} = - 1.448 (s_{1}-s_{2})
\label{murelative}
\end{equation}

Applying Eq.~(\ref{murelative}) to Coma, ABCG~85, and the group of NGC~4839, we
obtain:
\begin{eqnarray}
\mu_{\rm ABCG 85}-\mu_{\rm Coma} &=& 1.98 \pm 0.33\nonumber\\
\mu_{\rm NGC 4839}-\mu_{\rm Coma} &=& -0.51 \pm 0.53\,
\label{mudistances}
\end{eqnarray}

The mean entropy are taken from Table~\ref{stat}  while errors are
deduced from the standard error propagation.

In term of relative distances we have:
\begin{eqnarray}
D_{\rm ABCG 85} / D_{\rm Coma} &=& 2.5^{+0.4}_{-0.3} \nonumber\\
D_{\rm NGC 4839} / D_{\rm Coma} &=& 0.8^{+0.2}_{-0.5}\,
\label{distances}
\end{eqnarray}
The error bars seems especially large in the case of NGC~4839/Coma distance. Notice
however that the number of galaxies used in this case is very low (9 galaxies in the
group of NGC~4839); the error bars would be $\sim 0.2$ if we had a sample
of about a hundred galaxies in the group.

For comparison, the relative distance between ABCG~85 and Coma derived
from their mean redshifts is $\sim 2.4$, cf. Biviano et al. (1995b),
Durret et al. (1998). If we suppose that the distance of ABCG~85 is
given by (\ref{distances}) so that its redshift is a composition of
Hubble velocity plus its peculiar velocity, and \textit{a contrario}
that Coma is at rest in its Hubble frame, then ABCG~85 would have a
line of sight peculiar velocity of $\sim-600 \pm 240$km~s$^{-1}$.

\begin{figure}
	\centerline{\psfig{figure=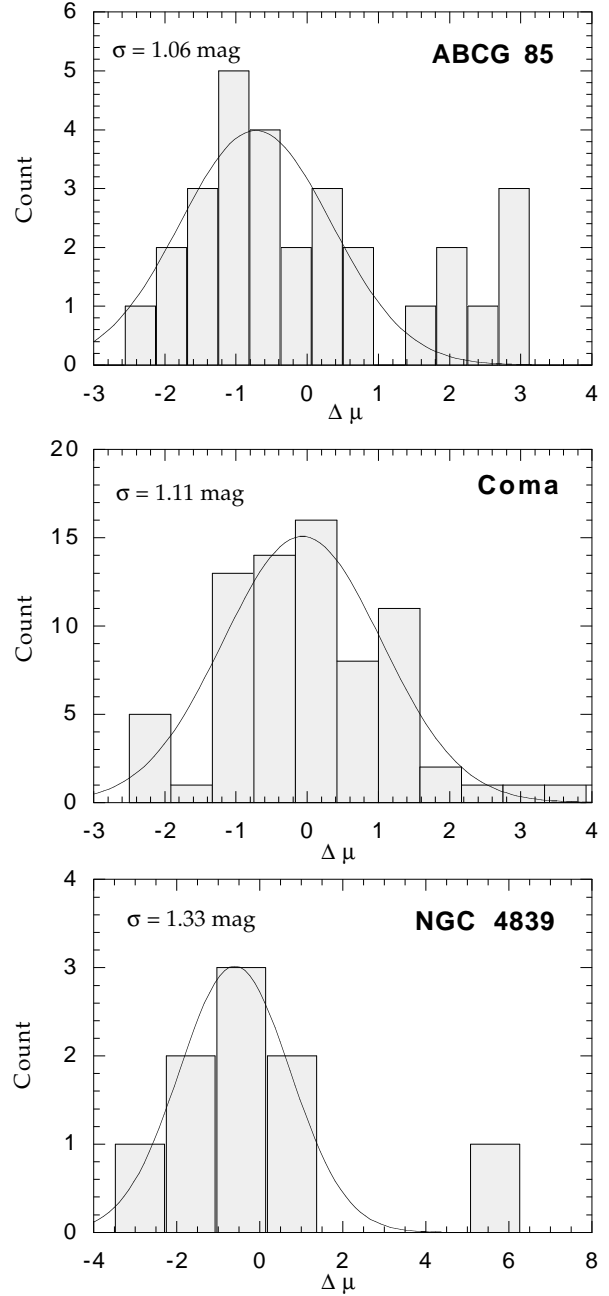,width=8cm}}
	\caption[]{Distribution of the distance 
	modulus, $\mu$, for each cluster computed using Eq.~(\ref{distRelative}). 
	The superposed curves are gaussian fits (using a standard least square 
	method) and in each panel is quoted the standard error in magnitude. Notice 
	that scales are different in each panel.}
	\label{fig:mu0}
\end{figure}

It seems that this method is not adequate to determine \textit{individual} 
distances inside a cluster. In Fig.~\ref{fig:mu0} we show for each cluster the 
distribution of the distance modulus, $\mu$, computed for each galaxy using the 
Eq.~(\ref{distRelative}) relative to the mean specific entropy (the values in 
Table \ref{stat}). The r.m.s. scatter is about 1.1 mag for the rich clusters.
Nonetheless, we note that using only the $a-\nu$ correlation, following 
Young \& Currie (1995) or Binggeli \& Jerjen (1998), and fitting it with an 
arbitrary polynomial of either 3 or 5 free parameters, we obtain scatters of the 
order of 0.83 mag.

\begin{figure}
	\centerline{\psfig{figure=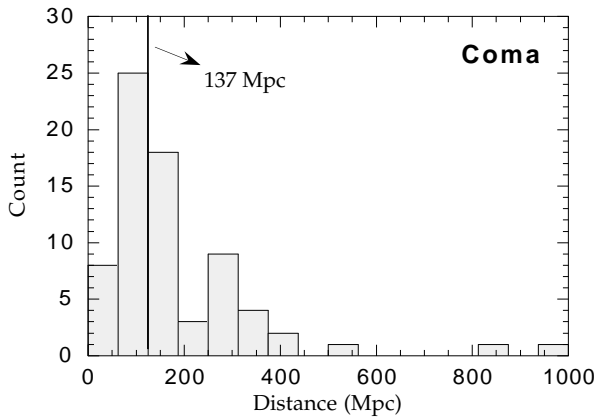,width=8cm}}
	\caption[]{Individual distances of the galaxies in Coma computed using
	Eq.~(\ref{distRelative}) and assuming the centre of the cluster is at
	137 $h^{-1}_{50}$~Mpc. Notice how this artificially stretches the
cluster in the line-of-sight direction.}
	\label{fig:distComa}
\end{figure}

For the Coma cluster we also show the dispersion in relative distances
(Fig.~\ref{fig:distComa}), assuming that the centre of Coma is at 137
$h^{-1}_{50}$~Mpc (Colless \& Dunn 1996; Durret et al. 1998).

The scatter around the Entropic Plane is not
produced by the distances of individual galaxies with respect to the
centre of mass, but it is due to a number of effects that will be
discussed in the next section. This scatter produces an artificially
elongated cluster; in the present case of Coma, we obtain a depth of
$\sim 1000$~Mpc or, neglecting the 5 most `distant' galaxies, a depth of $\sim 
350$~Mpc.

\section{Discussion}\label{discussion}

From previous work (Young \& Currie 1994, 1995; Binggeli \& Jerjen 1998) and
from our own data of elliptical galaxies in two rich clusters and a group,
it is clear that there are strong correlations among the primary free
parameters of the \Se\ profile. Before exploring the possible application
of this relationship, we tried to understand the physics underlying it.

Since isolated elliptical
galaxies seem to be systems in quasi-equilibrium, i.e., only  subject
to secular evolution, we assume that their dynamical state is
characterised by a stationary entropy. We can go further and postulate
that the entropy per mass unit (specific entropy, $s$) would be the
\textit{same} for all galaxies. In order to confront this hypothesis
against observational data, we needed to compute this specific
entropy. From
\Se 's light (and  mass) distribution for elliptical galaxies, and
assuming the equations of state of an ideal gas, as well as the isotropy of
the velocity
tensor, we have  derived a theoretical relationship  among the parameters
describing \Se 's profile.
This relation defines an `Entropic Plane'.

We have found that these primary  parameters for a
given cluster or group are well approximated by
the predicted constant specific entropy relation: elliptical galaxies seem
to share the same $s$.

We would like to emphasise that the two principal assumptions
that we have used,
i.e., the entropy and the ideal gas equations of state,
are not straightforward concepts in the context of stellar
self-gravitating dynamics. Interestingly, these supposedly inefficient tools
allow us to predict
relations which match quite well the observational data, a result which
raises interesting questions on the description of self-gravitating
systems.

Nevertheless, we note that the observations are quite dispersed
along the Entropy Line given by Eq.~(\ref{EntroLin}) and that there appears to 
be a tilt between de data and the predicted Entropy Line (Fig.~\ref{figEntroLin}).
A number of effects could account for this result.

\begin{itemize}
\item  Concerning the model:
\begin{itemize}
\item   we have used a
constant $M/L$ ratio. This is the simplest way to account for the relative
contribution of different light components in a galaxy, and for ellipticals
as a whole, given our knowledge of the $M/L$ trend in these galaxies
(J{\o}rgensen 1997).
Notice that the specific entropy varies as $\ln(M/L)$. Such a weak
dependence on $M/L$ (which we suppose constant from one galaxy to another),
may certainly explain why the predicted correlations work so well even if,
indeed, there is a variation of $M/L$ as a function of $\nu$ (and consequently
of the total mass of the galaxy);

\item for the hydrostatic equilibrium we have considered the
velocity dispersion tensor to be isotropic. This is probably
correct for galactic centres, but it is probably not for the outskirts.
At present, only the introduction of a phenomenological parameter
describing the variation of the
anisotropy with galactocentric distance would account for such a dispersion;

\item the equations of state that we have used are for ideal, non-interacting, 
hot gas. The inclusion of self-gravity explicitly (it is implicitly included by 
way of the hydrostatic equilibrium and Poisson equations) will probably tilt 
the Entropic Plane since the total mass of the galaxy correlates with the \Se\ 
parameter, in particular $\nu$;

\item  galaxies are likely to undergo merging and/or collisions, especially in
clusters; therefore during these  events they will necessarily escape from the
`Entropic Plane'. However, the resulting galaxies are expected to return
back to the `Entropic Plane' after a given relaxation time.

\end{itemize}

\item  Concerning the data analysis:
\begin{itemize}
\item In Fig.~(\ref{figNuDist}) we have given the distribution of $\nu$;
we note that this distribution is different from
that  of Virgo galaxies given by Graham \& Colless (1997). On the other hand,
our distribution of $\nu$ is similar to the one obtained by Binggeli \&
Jerjen (1998). This seems to
indicate that, in order to obtain comparable results, it is very
important to use the same procedure to analyse the data.
\end{itemize}
\end{itemize}

As noted in the previous section, galaxies in a cluster are not exactly at the
same distance from us, and
the scatter around the `Entropy line' is partly due to the various
(physical and observational) effects
noted above. The scatter is thus larger than that due only to
distance variations. We have used Formula~(\ref{distRelative}) for each galaxy
in the Coma cluster compared to the mean $s$ (the zero-point for the cluster).
The Coma cluster then appears extremely elongated along the
line-of-sight; Binggeli \& Jerjen (1998) conclude in the same vein when
discussing the application of the profile based distance indicator to
individual galaxies in the Virgo cluster made by Young \& Currie (1995)
[but see also Young \& Currie (1998), who refute the analysis of Binggeli \& Jerjen].

The `Entropic Plane' appears to be essentially useful to determine the \textit{distance
to clusters}, but not to individual galaxies inside a cluster, at least with the present scatter;
notice that it is also the case for the Tully-Fisher relation, which is used to
compare relative distances between clusters (Scodeggio et al. 1997).

\section{Conclusion}

The distance indicator we propose is not new, in the sense of using the distance 
independent shape parameter $\nu$ (or a function depending only of $\nu$) of the 
\Se\ law as a yardstick (cf. Young \& Currie 1994, 1995, who proposed a distance 
indicator based on the shape profile of elliptical galaxies and applied it to 
the Fornax and Virgo clusters). Our distance indicator, however, was derived 
from thermodynamics equilibrium, based on our \textit{strong} hypothesis: i.e. the 
specific entropy of relaxed galaxies is \textit{universal}. As noted in the 
introduction, this should be a natural consequence of some violent relaxation 
process; the universality, however, seems not to be guaranteed since the history 
or evolutionary stage of a given cluster may have some influence in the 
formation process of galaxies and consequently may produce some variations in 
the resulting specific entropy.

Therefore, seeking a physical explanation for the profile-shape 
distance indicators allowed us to derive a variant distance 
indicator that has only one free parameter, namely the specific entropy of 
galaxies. That should be compared with the empirical profile-shape
distance indicators [Young \& Currie 1994, 1995; Binggeli \& Jerjen (1998)] 
that need 3 to 5 free parameters. Nevertheless, since the exact shape of the
Entropic Plane depends on the way we model the elliptical galaxies, the 
empirical profile-shape distance indicators have a smaller intrinsic scatter.

Concerning the unique specific entropy of elliptical galaxies being independent
of the cluster in which they are located, if the existence of an universal 
profile (Navarro et al. 1995) is confirmed, the peculiar properties of clusters 
should not be very important. In this case, the relation expressed by 
Eq.~(\ref{EntroLin}) may provide some clue to the relaxation process in galaxies in addition to an attractive way to measure relative distances 
between clusters, using only photometrical data -- in contrast to the \FP\ which 
requires spectroscopic data as well.

\section{Acknowledgments}
We are very grateful to J.M. Alimi, H.V. Capelato and H. Verhagen for helpful
comments and discussions. We wish to thank F. Durret for making available the
ABCG~85 data to us and for helpful comments. We thank C. Lobo for helping us
with the Coma catalogue. This paper is based on observations collected at the
Canada-France-Hawaii telescope (operated by the National Research Council of
Canada, the \textit{Centre National de la Recherche Scientifique} of France,
and
the University of Hawaii) and the European Southern Observatory, La Silla,
Chile.

\section*{Appendix 1}

In Section \ref{thermo} we derive the specific entropy from the
fundamental relation of thermodynamics. Here, show that an equivalent
form is obtained using the Boltzmann-Gibbs entropy:
\begin{equation}
s = - \frac{S}{M} \int_{\Omega} F \ln(F) \diff^{3}x \diff^{3}v\, ,
\label{BoltzmannEntro}
\end{equation}
where $\Omega$ is the volume in phase space and $F$ is the
distribution function. From the definition of a mean value,
Eq.~(\ref{BoltzmannEntro}) can be re-written as:
\begin{equation}
s = \left\langle{\ln({F}^{-1})}\right\rangle \, .
\label{meanF}
\end{equation}

The distribution function can also be interpreted as the mass density in
phase space, $\diff M/(\diff^{3}x \diff^{3}v)$. Thus it is legitimate
to express $F$ as:
\begin{equation}
F \propto \rho/\sigma^{3}\, ,
\label{Fapprox}
\end{equation}
where $\sigma$ is the velocity dispersion and $\rho$ the density.
Now, we need an equation of
state. Adopting the following form:
\begin{equation}
P = K \rho^\gamma \sigma^{2}\, ,
\end{equation}
(a polytrope of index $n=1/(\gamma - 1)$ [Binney \& Tremaine 1987, p.~224])
and substituting it in Eq. (\ref{Fapprox}) and (\ref{meanF}) we finally get:
\begin{equation}
s = \left\langle{ \ln\left(
  \rho^{-(1+3\gamma/2)} P^{3/2} \right) }\right\rangle \, .
\end{equation}

For $\gamma = 1$ (ideal gas) we recover Eq.~(\ref{meanThermoEntro}).

\section*{Appendix 2}

The numerical computation of Eq.~(\ref{meanThermoEntro}) for an arbitrary
choice of $\nu$, $a$ and $\Sigma_0$ is cumbersome and slow.
Here, we will derive a simple analytical approximation for the specific entropy
for a galaxy following the \Se\ law. We do this in two steps.

\smallskip
\noindent (I) First we look for a simple functional form that describes $s$
as a function of the \Se\ parameters.

In section \ref{thermo} we show that the specific entropy can be
expressed as a mean value, cf. Eq. (\ref{meanThermoEntro}). We
rewrite it as:
\begin{equation}
s \approx -\frac{5}{2} \ln(\rho_{\rm eff}) + \frac{3}{2} \ln(P_{\rm eff})
+ C\, ,
\label{Entro_appendx}
\end{equation}
where we have used the approximation $\langle x\rangle \propto x_{\rm
eff}$, with
$x_{\rm eff}$ meaning the value of $x(r)$ taken at $r = R_{\rm eff}$, the
effective radius, and $C$ is a constant of proportionality. Although this
approximation
is justified for homologous, isotropic systems, it is in principle not so
for the \Se\
profile. Our goal here, however, is just to obtain a rough analytical form
that will later be fine tuned.

Both $\rho_{\rm eff}$ and $P_{\rm eff}$ can be computed numerically from
Eqs.~(\ref{reff2D}), (\ref{3Drho}) and(\ref{hydro}).
For the effective density, we have to a good approximation the following
analytical expression:
\begin{equation}
-\frac{5}{2} \ln(\rho_{\rm eff}) \approx -\frac{5}{2} \ln(\rho_0) +
4.75 \nu^{-1.46}\, ,
\label{roeff}
\end{equation}
and, for the effective pressure we have:
\begin{equation}
\frac{3}{2} \ln(P_{\rm eff})  \approx  \frac{3}{2}\ln(4\pi\;G) +
3\ln\left(\frac{a\;\rho_0}{\nu}\right) - 4.1\nu^{-1.0} \, .
\label{Peff}
\end{equation}
Both formulae are analytical fits that give the dependence of $\rho_{\rm
eff}$ and $P_{\rm eff}$ on the parameters of the \Se\ profile.

We also need the relation between the three dimensional central mass density,
$\rho_{0}$,
and the
projected central light density, $\Sigma_{0}$. This can be fairly well
approximated as:
\begin{equation}
\rho_{0} \approx \frac{\Sigma_{0}}{a} \; \frac{M}{L} \; 0.36\nu^{0.60}\, .
\label{ro0Sig0}
\end{equation}

Substituting Eqs. (\ref{roeff}), (\ref{Peff}), and (\ref{ro0Sig0}) into
Eq.~(\ref{Entro_appendx}), we obtain an analytical expression:
\begin{eqnarray}
s & \approx & \frac{1}{2}\ln(\Sigma_{0}) + \frac{5}{2}\ln(a) +
\frac{1}{2}\ln\left(\frac{M}{L}\right) + \frac{3}{2}\ln(G)\nonumber\\
	 & + & c_{0} + c_{1}\ln(\nu) - \frac{1}{\nu} + c_{2}\nu^{c_{3}}\, ,
\label{approxFin}
\end{eqnarray}
where the constants $c_{i}$ are to be determined and fine-tuned.

\smallskip
\noindent (II) The second step is to use the above formula as a template to
obtain an accurate approximation independent of $R_\eff$.

This is done by fitting the analytical
approximation, Eq.~(\ref{approxFin}), to the formula of the specific
entropy, Eq.~(\ref{meanThermoEntro}) (which can be computed numerically). We
then get the following values for the constants $c_{i}$:
\begin{displaymath}
c_0 = -0.299 ,\quad c_1 = -0.058 ,\quad c_2 = 3.656 ,\quad c_3 = -1.327 \, .
\nonumber
\end{displaymath}
The fit is done using a standard least-square method, computing 
Eq.~(\ref{meanThermoEntro}) in a `cube' with the axes $\nu$, $a$ and $\Sigma_0$. 
Taking the $M/L$ ratio as independent of the galactic radius as well as constant 
for all galaxies and scaling $G$ appropriately, we obtain the entropy relation 
expressed in Eq.~(\ref{Entro(a,nu,io)}), using the above numerical values for 
the $c_{i}$'s (the numerical value of $c_{0}$ has the term $(3/2)\ln G$ folded 
in).

The analytical approximation is accurate to better than 2\% in 
the range $0.15 \le \nu \le 1.85$ and the ratios $a_{\rm max}/a_{\rm min} = 
1.2\times 10^4$ and ${\Sigma_0}_{\rm max}/{\Sigma_0}_{\rm min} = 2.3\times 10^7$ 
(for $a$ and $\Sigma_0$ only the ratio is relevant since the values can be 
scaled with the choice of units of $G$). Considering that the $M/L$ ratio varies 
at most as $M^{0.2}$ (Pahre et al. 1995) and that the range in galaxy masses we 
are interested in is less than $10^4$, then the $M/L$ ratio varies at most by a 
factor $\sim 6$.

Although the functional dependence of the \Se\ parameters in 
Eq.~(\ref{approxFin}) was found with the help of the rough approximations given 
by Eqs.~(\ref{roeff}) and (\ref{Peff}), the $\Sigma_0$ and $a$ dependencies are 
the same as in the case of a sphere obeying the Plummer density profile (where 
we can calculate exactly the specific entropy).

\label{lastpage}
\end{document}